\documentclass{article} %
\usepackage{iclr2027_conference,times}

\usepackage{amsmath,amsfonts,bm}

\def\eqref#1{equation~\ref{#1}}

\def\1{\bm{1}}

\DeclareMathAlphabet{\mathsfit}{\encodingdefault}{\sfdefault}{m}{sl}
\SetMathAlphabet{\mathsfit}{bold}{\encodingdefault}{\sfdefault}{bx}{n}

\newcommand{\R}{\mathbb{R}}

 \usepackage{booktabs,pifont}
\usepackage{hyperref}
\usepackage{url}
\usepackage{algorithm}
\usepackage{algorithmic}
\usepackage{amsthm}
\usepackage{thmtools}
\usepackage{thm-restate}
\usepackage{graphicx}
\usepackage{tikz}
\usepackage{pgfplots}
\usepackage{caption}
\usepackage{hyperref} 
\pgfplotsset{compat=1.16}

\usetikzlibrary{positioning,calc,fit,arrows.meta}
\usetikzlibrary{backgrounds, positioning, calc, arrows.meta}
\usepackage{booktabs,amssymb,multirow}
\usepackage{wrapfig}
\usepackage{comment}
\usepackage{algorithm}
\usepackage{algorithmic}
\newcommand{\cmark}{\checkmark}\newcommand{\xmark}{$\times$}
\newtheorem{definition}{Definition}

\newtheorem{lemma}{Lemma}

\usepackage{xspace}
\makeatletter
\DeclareRobustCommand\onedot{\futurelet\@let@token\@onedot}
\def\@onedot{\ifx\@let@token.\else.\null\fi\xspace}

\makeatother

\newcommand{\Post}{\operatorname{Post}}
\newcommand{\Tube}{\operatorname{Tube}}
\newcommand{\Succ}{\operatorname{Succ}}
\newcommand{\Reach}{\operatorname{Reach}}
\newcommand{\overlinePost}{\overline{\Post}}
\newcommand{\overlineTube}{\overline{\Tube}}
\newcommand{\overlineReach}{\overline{\Reach}}
\newcommand{\cont}{\mathrm{cont}}

\newcommand{\method}{ReachECO\xspace}

\usepackage{xcolor}

\definecolor{istaDark}{HTML}{005844}

\pgfplotscreateplotcyclelist{myBoxplotColors}{
  {fill=blue!70!black},
  {fill=orange!85!black},
  {fill=red!75!black},
  {fill=violet!80!black},
  {fill=teal!70!black},
  {fill=magenta!70!black},
  {fill=gray!70!black},
  {fill=brown!70!black}
}
\title{Verification-Guided Safety-Repair of Learned Controllers}

\newcommand{\authorname}[1]{%
  \makebox[\dimexpr0.5\textwidth-2\tabcolsep\relax][l]{\textbf{#1}}}
\newcommand{\authoremail}[1]{{\fontsize{8.5}{10}\selectfont\texttt{#1}}}
\author{%
\authorname{Fabian Kresse}\\
ISTA \\
Klosterneuburg, Austria \\
\authoremail{fabian.kresse@ist.ac.at}%
\And
\authorname{Sobhan Gharanfoli}\\
University of Klagenfurt \\
Klagenfurt, Austria \\
\authoremail{sobhan.gharanfoli@aau.at}%
\AND
\authorname{Nikolaus Vertovec}\\
University of Oxford \\
Oxford, United Kingdom \\
\authoremail{nikolaus.vertovec@st-hughs.ox.ac.uk}%
\And
\authorname{Alessandro Abate}\\
University of Oxford \\
Oxford, United Kingdom \\
\authoremail{alessandro.abate@cs.ox.ac.uk}%
\AND
\authorname{Thomas A. Henzinger}\\
ISTA \\
Klosterneuburg, Austria \\
\authoremail{tah@ist.ac.at}%
\And
\authorname{Christoph H. Lampert}\\
ISTA \\
Klosterneuburg, Austria \\
\authoremail{chl@ist.ac.at}%
}

\iclrfinalcopy %
\begin{document}

\maketitle

\begin{abstract}
Formal verification can certify that a learned controller is safe. But what should 
be done when verification fails? 
Simply \emph{suppressing actions} that lead to unsafe next states is insufficient because 
the dangerous action that ultimately caused the safety violation might have occurred many 
steps earlier.
\emph{Retraining} modifies the policy globally, thereby potentially introducing new 
failures. %
Instead, we introduce \emph{\method}, a verification-guided repair procedure that exploits 
the information of failed multi-step reachability analysis for synthesizing explicit 
constraints on controller behavior. 
For Differentiable Weightless Controllers, the repairs compile into exact circuit 
patches that leave the controller unchanged everywhere outside the targeted policy cells. 
We prove  \emph{soundness} and  \emph{termination} of the repair procedure for \method, as well as, assuming 
exhaustive continuation search,  \emph{completeness} relative to the reachability abstraction. 
Empirically, we demonstrate \method's ability to successfully repair otherwise unsafe controllers 
in three continuous-control benchmarks: ACC, Docking2D and Car2D.
\end{abstract}

\section{Introduction}
Deep reinforcement learning (RL) has been extensively used to train agents that interact with digital environments, such as computer games~\citep{vinyals2019grandmaster}.  Recently, RL has also become an important part of post-training for large language models~\citep{ouyang2022training}.
The original inspiration for RL, however, was to synthesize controllers for agents that interact with the physical world, such as cyber-physical, robotic, and autonomous systems~\citep{rupprecht2022survey,tang2025deep}. 
Training these with RL is especially appealing when their desired behavior is difficult to hardcode, but can be expressed through a reward function~\citep{sutton1998reinforcement}.

While powerful, using RL for training systems that are meant to interact in real-world environments also has some downsides. 
In particular, empirically optimizing the reward function does not provide any formal \emph{guarantees} on the resulting controller~\citep{alshiekh2018safe, brunke2022safe, krasowski2022provably}. 
This is particularly consequential in safety-critical applications, such as cyber-physical systems, where failures can have severe physical consequences. 
Stating safety requirements is often straightforward. For example, an autonomous vehicle should not collide with another road user, a robotic system should not enter a forbidden region, or a physical process should remain within prescribed operating limits~\citep{zhao2023state}. However, \emph{enforcing} them is far from straightforward, as the consequences of an action may only become apparent several steps later. Arguing about safety therefore requires \emph{closed-loop} reasoning about the evolution of the system (controller as well as environment) over time. 

A common approach to closed-loop verification of neural-network 
control systems is to compute the set 
of reachable states, or an over-approximation of it, by repeatedly 
propagating sets through the neural-network controller and the system dynamics~\citep{xiang2018reachability, ivanov2019verisig,ivanov2021verisig, dutta2019sherlock, tran2020nnv, huang2022polar, schilling2022verification, lopez2023nnv, sidrane2022overt, tian2023taming}. 
If the set of all reachable states does not intersect with the unsafe set, the controller can indeed be certified to be safe. 
What, however, can be done if the sets do intersect?
\textbf{This is the central question we address in this work: given a trained 
controller that cannot be certified to be safe, how can it be \emph{repaired} 
such that it subsequently passes closed-loop safety verification?}

Existing approaches to \emph{closed-loop controller repair} can be broadly categorized
into two categories: \emph{local} and \emph{trajectory-level}.
\emph{Local methods} derive safety repairs from single-step 
constraints or counterexamples
~\citep{yang2022neural, santa2022safe}. %
However, a state might be safe itself, but have no available action
that will keep it safe in the next state. For instance, a car approaching 
an obstacle may already be traveling too fast for braking to avoid a later 
collision.
Consequently, single-step repair shifts the difficulty of reasoning about the
system's temporal dynamics to the availability of additional safety information
that summarizes future behavior, such as a larger \emph{losing set}, i.e., a
set of states from which an unsafe state is eventually unavoidable.
Computing such sets exactly, however, can be computationally intractable and is
even formally undecidable for sufficiently general hybrid systems~\citep{henzinger1995s}. %

\emph{Trajectory-level methods} influence the multi-step behavior of the 
controller, such that it avoids not only unsafe regions of the state 
space but also regions that inevitably lead to unsafe states at 
a later step~\citep{zhou2020runtime,hamilton2022training, lu2023causal, lyu2023autorepair, lu2025repairing}.
Such methods typically consider individual trajectories and do not consider the verifier 
results during the repair. Hence, they can only establish post-hoc formal guarantees
for \emph{all} possible trajectories. Furthermore, they generally rely on fine-tuning 
to repair the original neural network, thereby forfeiting any guarantees of termination 
or progress~\citep{boetius2023robust}. One exception is~\citet{lu2023causal}, which searches 
directly over counterfactual controller behaviors. Its guarantee, however, is tied to the analyzed run-time violation
rather than to all trajectories from an initial-state region. %

In this work, we take inspiration from the setting of static feed-forward networks. 
There, neural-network repair can be formulated directly as a problem of enforcing 
explicit constraints on the output: verification may 
identify inputs or regions whose 
outputs must change, and the repaired network is required to satisfy these constraints. 
This approach enables strong formal guarantees on the repair procedure~\citep{goldberger2020minimal,sotoudeh2021provable,
fu2022reassure,tao2023architecture}.
However, extending this idea to closed-loop systems is nontrivial because the required 
modification may lie several time steps before the observed safety violation, and the repair 
constraint must account for the resulting future behavior.

\textbf{Our main contribution, \method, solves this problem in an effective yet efficient
way for abstraction-based controllers. \method uses closed-loop reachability analysis to 
derive repair constraints over future behavior, bringing constraint-based repair and 
its formal guarantees to the multi-step closed-loop setting.}

Specifically, we formulate \method for \emph{abstraction-based neural-network controllers}
~\citep{jin2022trainify,tian2023taming,tian2023boosting} and instantiate it with
Differentiable Weightless Controllers (DWCs)~\citep{kresse2026differentiable}: 
these are controllers that have an exact boolean circuit representation, such that repairs 
can be formulated as exact local circuit patches. We then prove termination and, for an 
exhaustive variant, completeness relative to the environment over-approximation. Finally 
we evaluate \method on several closed-loop control benchmarks.

\section{Verification-Guided Closed-Loop Repair}
We now present \textbf{Reach}ability-derived \textbf{E}ngineering \textbf{C}hange \textbf{O}rder (ReachECO), a verification-guided procedure that derives controller-repair constraints from failed multi-step closed-loop reachability analysis. The name is inspired by engineering change orders (ECOs), which modify an existing circuit implementation to satisfy revised design requirements while preserving as much of the original implementation as possible~\citep{lin1995logic,krishnaswamy2009deltasyn}. Similarly, \method seeks to restore certifiability through localized patches to the controller rather than replacing it globally. %
Before we do so, we introduce the main concepts necessary to reason about the properties of discrete systems that operate in a continuous environment, in particular their safety. Formal definitions and additional details are given in Appendix~\ref{sec:formal}.

\paragraph{Abstraction-based Control.} 
We study a control task with continuous state space $\mathcal{X}\subseteq\R^d$ and system dynamics governed by a 
differential equation, $\dot x = f(x,a)$, where $a\in\mathcal{A}$ 
is a control action. 
We consider controllers $\pi:\mathcal X\to\mathcal A$ that are evaluated
every $\Delta$ time units, with the selected action held constant between
controller updates (called \emph{zero-order hold}). 
While cyber-physical systems necessarily have to process real-valued inputs, because 
they operate in continuous (real-world) environments, it has been observed that forming 
\emph{abstractions} of the continuous state-space into finite sets can often preserve 
near-optimal behavior~\citep{abel2019theory}. 
Similarly, it has been observed that strong \emph{quantization} of a neural controller's 
input space often incurs only limited performance degradation for standard continuous-control tasks~\citep{kresse2025learning,kresse2026differentiable}. 

Such abstractions can substantially simplify formal verification: controllers defined over a finite input partition avoid much of the complexity associated with having to bound neural network activations over continuous input regions, enabling significantly more scalable closed-loop reachability analysis~\citep{jin2022trainify,tian2023boosting, tian2023taming,yang2025formal}. 
Consequently, in this work, we focus on \emph{piecewise-constant abstraction-based controllers} 
(Def. \ref{def:pcabc}, Appendix \ref{sec:formal}), which partition the state space into finitely
many axis-aligned \emph{policy cells} $\mathcal Q$ and assign a single action $a\in\mathcal A$ to each cell $q\in \mathcal Q$.

\paragraph{Differentiable Weightless Controllers.}
Importantly to maintain efficiency, the use of abstraction-based controllers 
is not contradictory to gradient-based learning. In particular, 
\emph{Differentiable Weightless Controllers (DWCs)}~\citep{kresse2026differentiable} 
can be trained with gradient descent but nevertheless admit an exact 
representation as a Boolean circuit. Their quantized input encoding 
creates the finite policy partition $\mathcal Q$, while their quantized 
outputs lead to a finite action alphabet $\mathcal A$. Hence, DWCs provide 
a natural instance of the abstraction-based controller class considered here. 
DWC circuits consist primarily of Boolean lookup tables (LUTs). A $6$-input
LUT (LUT6), as commonly provided by modern FPGAs, stores the truth table of an
arbitrary Boolean function of six inputs, requiring only $2^6=64$ bits~\citep{xilinx_lut6}. 
Consequently, DWC inference can be implemented directly as Boolean 
circuit evaluation without floating-point multiplication.
Despite this strongly quantized representation, DWCs have been shown to learn policies 
of comparable quality as more conventional neural-network controllers on challenging 
continuous-control tasks, such as Humanoid~\citep{kresse2026differentiable}. 
Hence, we instantiate our framework for DWCs. 
In Section~\ref{sec:patch-circuit}, we exploit the exact circuit representation to realize 
repairs as bit-exact local patches of the original controller. %

\paragraph{Abstraction-Based Closed-Loop Reachability Analysis.}
\emph{Closed-loop reachability analysis} propagates sets of possible states through 
the controller and system dynamics. Given a set $B\subseteq\mathcal X$ and
action $a$, we assume \emph{sound} over-approximations $\overlinePost(B,a)$ of the 
one-step successor set and $\overlineTube(B,a)$ of all states visited before the 
next controller step, where \emph{soundness} means that no trajectory is 
omitted that the true system starting in $B$ under $a$ could take. 
Every over-approximation is further enclosed by an axis-aligned hyperrectangle 
in $\mathbb R^d$, which we call a \emph{box}, to enable efficient propagation 
by the verifier that performs the reachability analysis.

For an abstraction-based controller, let every box $B$ be contained in exactly 
one policy cell, which we denote by $q(B)\in\mathcal Q$. Hence, the controller
assigns a single exact action $\pi(q(B))$ to all states in $B$.
For such a box, we  denote by $\Succ(B,a)$ a set of \emph{successor boxes} by 
computing $\overlinePost(B,a)$ and, because the post-image may intersect multiple 
of the policy cells, subsequently splitting along policy-cell boundaries. 
For scalability, if multiple boxes fall into the same policy cell, they 
may also be merged again~\citep{jin2022trainify,tian2023boosting, tian2023taming, yang2025formal}.

Now, starting from an initial set of boxes, $\mathcal{X}_0$, repeated 
application of $\Succ$ yields a sound over-approximation of the closed-loop reachable set. 
Let $\overline{\mathcal X}_k$ denote the collection of reachable boxes at step $k$, 
then we define, for any time horizon $T\in\mathbb{N}$:
\begin{equation}
\overlineReach_{\pi,T}(\mathcal X_0)
:=
\bigcup_{k=0}^{T-1}
\bigcup_{B\in\overline{\mathcal X}_k}
\overlineTube(B,\pi(q(B))).
\end{equation}

Consequently, $\overlineReach_{\pi,T}(\mathcal X_0)$ over-approximates every state visited by the 
closed-loop system from $\mathcal X_0$ during the first $T$ control steps.
We now have all components to formally define our central object of interest: \emph{certified safety}.
\begin{definition}[Certified safety]\label{def:certified-safety}
A controller $\pi$ is called \emph{certified safe to horizon $H\in\mathbb{N}$}, if %
\begin{equation}
    \overlineReach_{\pi,H}(\mathcal X_0) \cap \mathcal U = \emptyset, \label{eq:certifiedsafety}
\end{equation}
{where $\mathcal{U}\subseteq\mathcal{X}$ is the set of unsafe states.}
We call $\pi$ \emph{certified safe to horizon $H=\infty$} %
if for some step $k$ the reachable-set computation establishes an inductive fixed point,
\begin{equation}
    \textstyle\bigcup
    \overline{\mathcal X}_{k+1}
    \subseteq
    \bigcup
    \overline{\mathcal X}_k,
    \qquad\text{and}\qquad
    \overlineReach_{\pi,k+1}(\mathcal X_0)\cap\mathcal U=\emptyset. %
\end{equation}
\end{definition}

In addition, for a reach-avoid task, any trajectory that has entered the absorbing goal region is subsequently assumed to be safe.

A controller that has been proven \emph{certified safe} is also safe under the actual dynamics $f$. Conversely, however, the inability to certify a controller does not imply that it is necessarily unsafe.
This, however, is not a shortcoming of this specific definition. It is known that \emph{exact reachability}, 
and therefore safety, is not only computationally intractable but formally \emph{undecidable} for sufficiently 
general hybrid systems~\citep{henzinger1995s}.

\paragraph{Controller Repair.}\label{sec:repair}
When a controller fails safety verification, a 
natural next step is to \emph{repair} it. 
Formally,  we define a \emph{repair instance} as an abstraction-based 
controller $\pi_0$ together with sound propagation operators $\overlinePost$ 
and $\overlineTube$ for the system dynamics, an initial set $\mathcal X_0\subseteq\mathcal X$, an unsafe set $\mathcal U\subseteq\mathcal X$, 
an optional absorbing goal set $\mathcal {X_G}\subseteq\mathcal X$, and 
a verification budget $T$. 
The goal is to obtain a controller $\pi'$ that is \emph{certified safe}
in the sense of Definition~\ref{def:certified-safety}, while retaining 
as much of the behavior of the original controller $\pi_0$ as possible. 
Repairs may assign each policy cell a new action from a finite repair alphabet $a\in \mathcal{A}_{cont} \subseteq \mathcal A$ or keep the original action $\pi_0(q)$.

\subsection{\method}\label{sec:circuit-repair}

\begin{figure}[t]
\begin{minipage}[t]{0.485\textwidth}
\begin{algorithm}[H]
\caption{\textsc{\method} %
}
\label{alg:lazy-repair}
\begin{algorithmic}[1]
\STATE \textbf{Input:} $\pi$, $\mathcal X_0$, $\underline{\mathcal L}$,
       budget $T$, backend $\mathcal W$

\STATE $\pi'\gets\pi$;\;
       $\Gamma\gets\emptyset$;\;
       $\overline{\mathcal X}_0\gets\mathcal X_0$;\;
       $k\gets 0$

\WHILE{$k < T$}

\STATE $(\overline{\mathcal{X}}_{k+1},\mathcal F)\gets
    \textsc{ReachStep}(
        \overline{\mathcal{X}}_k,\pi',\underline{\mathcal L})$
        \label{line:reachstep}

    \IF{$\mathcal F \neq \emptyset$}
        \label{line:safecheck}

        \STATE $(s,\pi',\Delta\underline{\mathcal L})\gets
            \textsc{BRepair}(\mathcal F, \underline{\mathcal L}, \mathcal W, k)$
            \label{line:backwardrepair}

        \STATE $\underline{\mathcal L}\gets
            \underline{\mathcal L}\cup\Delta\underline{\mathcal L}$
            \label{line:expand}
            \COMMENT{certified losing}

        \IF{$s=\textsc{Unknown}$}
            \RETURN \textsc{Unknown}
        \ENDIF

        \STATE $k \gets
            \textsc{EarliestAffected}(\pi',\Delta\underline{\mathcal L})$
            \label{line:restart}

    \ELSIF{$\bigcup\overline{\mathcal X}_{k+1}\subseteq \bigcup\overline{\mathcal{X}}_k$}
        \label{line:fixedpoint}
        \RETURN $(\pi',\infty)$
        \COMMENT{inductive invariant}

    \ELSE
        \STATE $k \gets k+1$
    \ENDIF

\ENDWHILE
\RETURN $(\pi',T)$

\end{algorithmic}
\end{algorithm}
\end{minipage}\hfill
\begin{minipage}[t]{0.485\textwidth}
\begin{algorithm}[H]
\caption{\textsc{BRepair}. $\Gamma$ is the set of committed constraints and is stored across calls.}
\label{alg:backward-repair}
\begin{algorithmic}[1]

\STATE \textbf{Input:} next step failing boxes $\mathcal F, \underline{\mathcal L}, \mathcal W, t$
\STATE $\overline\Gamma\gets\emptyset$;\;
       $L_0\gets\underline{\mathcal L}$
\WHILE{\textbf{true}}
    \STATE $(\Gamma_w,\mathcal F_o,\Delta\underline{ \mathcal{L}})\gets
        \textsc{FindCont}(\mathcal F, \mathcal W)$
\STATE $\underline{\mathcal L}
    \gets \underline{\mathcal L}\cup\Delta\underline{\mathcal L}$;\;
    $\overline\Gamma\gets\overline\Gamma\cup\Gamma_w$
    \IF{$\mathcal F_o=\emptyset$}
        \STATE $(s,\pi',\mathcal F_o,\Gamma,\overline\Gamma)\gets
            \textsc{Joint}(\Gamma,\overline\Gamma)$
        \IF{$s=\textsc{Sat}$}
            \STATE $\Gamma\gets\Gamma\cup\overline\Gamma$
            \RETURN $(\textsc{Repair},\pi',
                \underline{\mathcal L}\setminus L_0)$
        \ELSIF{$s=\textsc{Unknown}$}
            \RETURN $(\textsc{Unknown},\pi',
                \underline{\mathcal L}\setminus L_0)$
        \ENDIF
    \ENDIF
    \IF{$\exists(B,h)\in\mathcal F_o:\ h=t+1$}
        \RETURN $(\textsc{Unknown},\pi',\underline{\mathcal{L}}\setminus L_0)$
    \ENDIF
    \STATE $\mathcal F \gets
        \textsc{Backtrack}(\mathcal F_o)$
\ENDWHILE
 
\end{algorithmic}
\end{algorithm}
\end{minipage}
\end{figure}

\method enables effective yet efficient controller repair.
Algorithms~\ref{alg:lazy-repair} and \ref{alg:backward-repair} provide pseudo-code, 
based on a number of arguments and subroutines that we will introduce in this 
section. Appendix~\ref{sec:formal} gives additional details and formal definitions. We provide a high-level sketch of \method's working principles 
and how it achieves local repair in a closed-loop setup.

At its core, \method alternates between forward closed-loop reachability
analysis and backward repair search. It maintains the sets of reachable boxes
$\overline{\mathcal X}_0,\ldots,\overline{\mathcal X}_k$ computed under the 
current controller together with a certified losing set $\underline{\mathcal L}$.
The latter contains regions for which the algorithm has established that
no admissible continuation can avoid a future safety violation.
Initially, $\underline{\mathcal L}$ may contain only the unsafe set
$\mathcal U$ and can be expanded as additional losing regions are certified. 
Throughout $\underline{\mathcal L}$ is a subset of the \emph{true} losing set $ \mathcal{L}^*$. We call action $a$ \emph{one-step safe} from a box $B$ if its
inter-sample tube avoids $\mathcal U$ and its post-image avoids the current
losing set underapproximation:
\begin{equation}
\label{eq:safe step}
    \mathrm{SafeStep}(B,a)
    \;:\Longleftrightarrow\;
    \overlineTube(B,a)\cap\mathcal U=\emptyset
    \;\land\;
    \overlinePost(B,a)\cap\underline{\mathcal L}=\emptyset .
\end{equation}

Suppose forward verification encounters one or more boxes
$B\in\overline{\mathcal X}_k$ for which $\mathrm{SafeStep}(B,\pi(q(B)))$ is not fulfilled.
\method collects these failing boxes as the initial \emph{repair obligations}
$\mathcal F$. We then search for alternative controller actions
that admit a safe continuation.
The continuation backend $\mathcal W$ determines which candidate actions are
explored. If no continuation exists from a failing box, then intervening at
that time step is already too late, and \method backtracks to
$\overline{\mathcal X}_{k-1}$. The required \emph{safe} continuation horizon, however,
increases by one as we want to establish a safe continuation up to $k+1$, precisely where the initial safety violation was encountered. This backtracking and forward search process continues until a
safe continuation is found for every failure or the initial set is reached.
During this search, \method may also certify additional regions as losing and
add them to $\underline{\mathcal L}$.

Each successful continuation search yields a repair \emph{constraint}. 
Each constraint ensures a safe continuation from the obligation that
generated it.
Constraints discovered across different failures and repair rounds
are accumulated and must ultimately be satisfied by a new
controller $\pi'$. This joint realization is nontrivial because independently derived
constraints may require incompatible actions in the same policy cell.
Section~\ref{sec:constraint-handling} describes how \method resolves such
conflicts while preserving progress of the repair procedure.
Once a jointly realizable repair is found, the controller is updated.
Because this change may alter previously computed reachable sets,
\method resumes forward verification from the earliest time step affected by
either the controller modification or an expansion of
$\underline{\mathcal L}$.

In summary, \textbf{at the core of \method lies an intuitive procedure of 
alternating forward propagation with lazy backtracking. 
Our main contribution lies in how safety failures are handled when they 
are encountered: resolving conflicts while ensuring finite progress 
and, under exhaustive-search assumptions, completeness relative to 
the verifier abstraction.}

\subsection{Obligations, Constraints and Conflicts}
\label{sec:constraint-handling}

We next describe how \method represents and combines the repairs discovered
during backward search. We first introduce \emph{obligations}, which encode
the need to establish a safe continuation from a reachable region. We then
show how successful continuation searches are stored as \emph{repair
constraints}. Finally, we describe how constraints from different failures
are jointly resolved while ensuring progress of the repair procedure.

\paragraph{Obligations.} An obligation is a pair $(B,h)$, where $B$ is a box and $h$ is the
number of future control steps for which a safe continuation must be
established. The initial obligation set $\mathcal F$ is obtained from the boxes that violated the \textsc{SafeStep} predicate during forward reachability 
propagation. Each one is initialized with horizon
$h=1$. Starting from $B$, \textsc{FindCont} uses the same sound set propagation as
forward verification, but explores alternative actions proposed by
$\mathcal W$. 

Thus, for an obligation $(B_0,h)$, the search has the following alternating structure:
\begin{equation}
\label{eq:safety game}
\begin{aligned}
\exists a_0 \;&:\; \mathrm{SafeStep}(B_0,a_0)
\\[-0.2em]
&\land\;
\forall B_1\in\Succ(B_0,a_0)\;
\exists a_1 \;:\; \mathrm{SafeStep}(B_1,a_1)
\\[-0.2em]
&\hspace*{.25\textwidth}\vdots
\\[-0.2em]
&\land
\forall B_{h-1}\in\Succ(B_{h-2},a_{h-2})\;
\exists a_{h-1} \;:\; \mathrm{SafeStep}(B_{h-1},a_{h-1}).
\end{aligned}
\end{equation}
The backend $\mathcal W$ determines which candidate actions are explored. An
exhaustive backend considers every action in $\mathcal A_{\mathrm{cont}}$. This leads 
to a finite-horizon safety game-like search~\citep{tabuada2009verification}.

If \textsc{FindCont} cannot establish an
$h$-step safe continuation from $B$, the obligation is returned as unresolved ($\mathcal F_o$).
\textsc{BackTrack} then replaces it by obligations on its predecessor box at 
the preceding time step and increases the required
continuation horizon to $h+1$. Thus, every backtracking step searches for an
earlier intervention while accounting for one additional future step. If no safe continuation can be established from time step $0$, \method returns \textsc{Unknown}: under the employed reachability abstraction and continuation
backend, no feasible continuation could be established from that obligation.

\paragraph{Constraints.} 
A successful continuation search is represented by a finite game DAG
$\mathcal G$ containing the safe continuations discovered by the backend
$\mathcal W$.
Each node $v$ corresponds to a box $B_v$ and stores the candidate actions
that admit a safe continuation for the remaining horizon $h_v$.
At the root node $v_0$, these are exactly the box and horizon of the
corresponding obligation. The successor boxes generated by each 
action that was established to satisfy Equation~\ref{eq:safety game},
appear as child nodes with horizon $h_v-1$. 

Together with its root time step $i$, the game forms a
\emph{repair constraint}
\(
    \gamma=(\mathcal G,i).
\)
By retaining all feasible continuations found by $\mathcal W$, rather than
committing immediately to a single one, the subsequent joint realization step
has flexibility when combining repairs from different failures.
We call $\gamma$ \emph{anchored} if $i=0$, since its root has reached the
initial time step and cannot be backtracked further.  \method maintains two sets of constraints. The \emph{committed} constraints
$\Gamma$ were established in previous calls to \textsc{BRepair}, whereas the
\emph{current} constraints $\overline{\Gamma}$ were generated during the
present call. This distinction matters during conflict resolution because
committed constraints were generated under an earlier controller.

\paragraph{Conflict Resolution.} Once every current obligation has produced a repair constraint,
\textsc{Joint} searches for a single stationary controller that satisfies all
accumulated constraints. Satisfying a constraint means selecting one of its continuations: 
whenever a game node is reached, the controller must
choose one of the actions stored at that node. 
Since every box belongs to a policy cell $q\in\mathcal Q$,
realizing a continuation ultimately imposes action requirements on these
cells.

Constraints generated independently need not be jointly realizable. In
particular, different continuations may require incompatible actions in the
same policy cell, while the repaired controller must assign exactly one
action to each cell. We encode this finite joint-realization problem using
CP-SAT~\citep{cpsatlp}, requiring one action per policy cell and satisfaction
of every repair constraint. Among feasible assignments, we prefer
controllers that remain close to the original policy.

Accordingly, \textsc{Joint} has three possible outcomes:
\textsc{SAT}, returning a jointly satisfying controller;
\textsc{Backtrack}, returning a set of conflicting non-anchored constraints
as new obligations $\mathcal F_o$; or \textsc{Unknown}, if the conflict
cannot be resolved because it involves only anchored constraints. If 
the constraints are jointly infeasible, \method first backtracks
conflicting non-anchored constraints generated in the current repair round. 
If the conflict instead involves
\emph{anchored} constraints, non-anchored committed constraints may be
discarded in favor of newly anchored constraints. Anchored
constraints are never discarded. 
Consequently, \method returns \textsc{Unknown} only when repair fails at the
initial time step: either no safe continuation can be established for an
anchored obligation, or the anchored constraints are jointly
infeasible under the employed abstraction and backend.

\newpage
\subsection{ECO Patch Compilation}
\label{sec:patch-circuit}
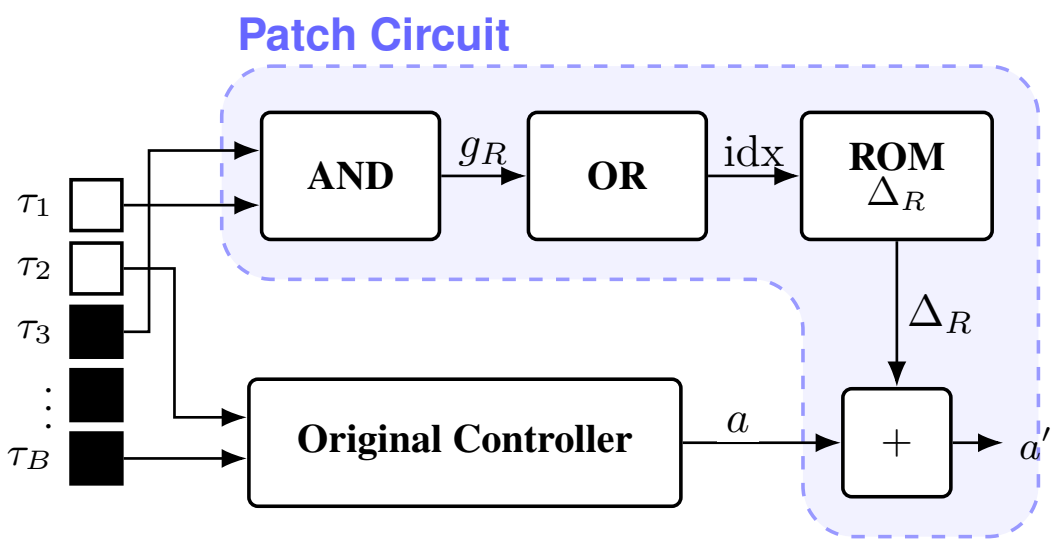
\begin{wrapfigure}[13]{r}{0.52\textwidth}
\vspace{-1.6cm}
    \centering
\resizebox{\linewidth}{!}{%

\pgfdeclarelayer{background}
\pgfsetlayers{background,main}

\begin{tikzpicture}[
    >=Latex,
    font=\small,
    circuitbox/.style={
        draw, thick, rounded corners=2pt, minimum height=1.0cm, minimum width=1.4cm, align=center,
        fill=white
    },
    plainbox/.style={
        draw, thick, rounded corners=2pt, minimum height=1.0cm, minimum width=1.5cm,
        align=center, fill=white
    },
    therm/.style={draw, thick, minimum width=4mm, minimum height=4mm, inner sep=0pt},
    wire/.style={->, semithick},
    buslabel/.style={font=\normalsize, inner sep=2pt, fill=white},
    patchlabel/.style={font=\normalsize, inner sep=2pt, fill=none},
]

\node[therm, fill=white, label=left:{$\tau_1$}] (t1) at (0, 0.72) {};
\node[therm, fill=white, label=left:{$\tau_2$}, below=2pt of t1] (t2) {};
\node[therm, fill=black, label=left:{$\tau_3$}, below=2pt of t2] (t3) {};
\node[therm, fill=black, label=left:{$\vdots$}, below=2pt of t3] (t4) {};
\node[therm, fill=black, label=left:{$\tau_B$}, below=2pt of t4] (t5) {};

\node[circuitbox] (and)  at (2.0,  0.95) {\textbf{AND}};
\node[circuitbox] (or)   at (4.1,  0.95) {\textbf{OR}};
\node[plainbox]   (bram) at (6.3,  0.95) {\textbf{ROM}\\[-2pt]{\footnotesize$\Delta_R$}};

\node[circuitbox, minimum width=3.4cm] (orig) at (2.9, -1.15) {\textbf{Original Controller}};
\node[plainbox, minimum width=0.85cm, minimum height=0.85cm] (add) at (6.3, -1.15) {$+$};
\node[right=0.4cm of add] (out) {$a'$};

\draw[wire] (t1.east) -- (and.west  |- t1);
\draw[wire] (t5.east) -- (orig.west |- t5);
\draw[wire] (t3.east) -- ++(0.2,0) |- ($(and.west)+(0,0.2)$);
\draw[wire] (t2.east) -- ++(0.4,0) |- ($(orig.west)+(0, 0.2)$);

\draw[wire] (and.east) -- node[patchlabel, above]{$g_R$} (or.west);
\draw[wire] (or.east)  -- node[patchlabel, above]{$\mathrm{idx}$} (bram.west);

\draw[wire] (bram.south) -- node[patchlabel, right]{$\Delta_R$} (add.north);
\draw[wire] (orig.east)  -- node[buslabel, above, pos=0.35]{$a$} (add.west);
\draw[wire] (add.east) -- (out.west);

\begin{pgfonlayer}{background}
    \coordinate (C1) at ($(and.north west)+(-0.3, 0.35)$);

    \coordinate (tmp2) at ($(bram.east)+(0.35, 0)$);
    \coordinate (C2) at (tmp2 |- C1);

    \coordinate (tmp3) at ($(add.south)+(0, -0.3)$);
    \coordinate (C3) at (C2 |- tmp3);

    \coordinate (tmp4) at ($(add.west)+(-0.3, 0)$);
    \coordinate (C4) at (tmp4 |- C3);

    \coordinate (tmp5) at ($(bram.south)+(0, -0.3)$);
    \coordinate (C5) at (C4 |- tmp5);

    \coordinate (C6) at (C1 |- C5);

    \draw[fill=blue!5, draw=blue!40, dashed, thick, rounded corners=4mm]
        (C1) -- (C2) -- (C3) -- (C4) -- (C5) -- (C6) -- cycle;

    \node[font=\sffamily\bfseries\color{blue!60}, anchor=south west]
        at (C1) {Patch Circuit};
\end{pgfonlayer}

\end{tikzpicture}}
\caption{ECO patch architecture. Existing thermometer predicates are AND-reduced to implement
the policy-cell identification $g_R$, from these an address selecting the
stored correction $\Delta_R$ is obtained. The correction is applied exactly and only to repaired policy cells.}
\label{fig:patch-circuit}
\end{wrapfigure}

\label{sec:patch-compilation}
The repair from \textsc{Joint} returns a set of policy cells $\mathcal Q_r \subseteq \mathcal Q$, where actions differ from the original controller. The repair is realized as:
\begin{equation}
\label{eq:patch}
\pi'(x)
=
\pi_0(x)
+
\sum_{q_r\in\mathcal Q_r}
g_{q_r}(x)\,\Delta_r,
\end{equation}
For DWCs, an initial thermometer encoding~\citep{carneiro2015multilingual} provides the threshold predicates
needed to identify each policy cell. We therefore implement each $g_{q_r}$ as
an AND reduction over these bits. The resulting one-hot guards are encoded into
a memory address selecting the corresponding correction $\Delta_r$, which is
added to the DWC's integer output. Address zero stores the zero correction, so
the controller is unchanged outside the repaired cells. The final BRAM-based
output decoding stage of the DWC remains unchanged. Figure~\ref{fig:patch-circuit} illustrates the constructed patch. 
It is similar to
support-network repair in REASSURE~\citep{fu2022reassure} and lookup-based
patch architectures~\citep{tao2023architecture}, but realizes the repair
exactly as a digital circuit. Note that the patch can be executed largely in parallel with the original controller. 
Circuit synthesis and implementation
details are given in Appendix~\ref{app:dwc-size}.

\subsection{Theoretical Guarantees}
\label{sec:theory}

The purpose of \method is to provide controllers with formal guarantees; hence,
the repair procedure itself should also admit formal guarantees.
We state these below as theorems. The corresponding proofs and implementation
assumptions are given in Appendix~\ref{app:proofs}.
Soundness applies to the full procedure. Completeness is stated with respect to the final learned losing set of \method.

\begin{restatable}[Soundness]{theorem}{thmsoundness}
\label{thm:soundness}
Suppose ReachECO returns $(\pi', H)$ with $H \in \{T,\infty\}$ rather than
\textsc{Unknown}. Then $\pi'$ is certified safe to horizon $H$ (Definition~\ref{def:certified-safety}).
In particular, no trajectory of the closed-loop system under $\pi'$ starting in
$\mathcal{X}_0$ enters $\mathcal{U}$ during the first $H$ control steps, and for
all time if $H=\infty$.
\end{restatable}

\begin{restatable}[Termination]{theorem}{thmtermination}
\label{thm:termination}
Assume that the initial set has a finite abstract representation, and that the
policy partition $\mathcal Q$, the continuation alphabet $\mathcal A_{\cont}$,
and the verification horizon $T$ are finite. Furthermore, assume \textsc{Joint}
always returns a repair if there exists one under the constraints.
Consider monotone-losing set expansion and persistent merge-mask refinement as described in Appendix \ref{sec:formal}.
Then \method terminates and returns either
$(\pi',H)$, with $H\in\{T,\infty\}$, or \textsc{Unknown}.
\end{restatable}

Here $\mathrm{Safe}^{L'}_T(B,\pi)$ denotes that every transition in the
unmerged $T$-step unfolding from $B$ under $\pi$ satisfies
$\mathrm{SafeStep}$ with respect to a losing set $L'$. This losing set is the last losing set constructed during execution of \method and contains $\mathcal U$. The recursive definition is given in
Appendix~\ref{sec:formal}. We denote with $\mathcal A^{\mathcal Q}$ the set of all controllers with the action alphabet $\mathcal A$ over the grid $\mathcal Q$.

\begin{restatable}[Completeness under exhaustive search]{theorem}{thmcompleteness}
\label{thm:relative-completeness}
Under the assumptions of Theorem~\ref{thm:termination}, let $\mathcal X_0$ be
the fixed initial box representation, containing every anchored root. Let $L'$ denote the final losing set constructed during execution. Use
the exhaustive backend over the original action alphabet $\mathcal A$.
If some $\overline\pi\in\mathcal A^{\mathcal Q}$ satisfies
\[
    \forall B\in\mathcal X_0:\quad\mathrm{Safe}^{L'}_T(B,\overline\pi),
\]
then \method returns a controller $\pi'\in\mathcal A^{\mathcal Q}$ that is
certified safe (Definition~\ref{def:certified-safety}). %
\end{restatable}

While exhaustive continuation search is largely aspirational, due to potentially large action spaces, Corollary \ref{cor:backend-completeness} follows immediately from it. Importantly, Corollary \ref{cor:backend-completeness} implies that \method always returns a repair if there exists one in the searched action alphabet and under the utilized Tube and Post image overapproximations and \textsc{Joint} does not time out.

\begin{restatable}[Completeness relative to the continuation backend]
    {corollary}{corbackendcompleteness}
\label{cor:backend-completeness}
Under the assumptions of Theorem~\ref{thm:termination},
suppose there exists a stationary controller
$\overline\pi\in\Pi_R$ satisfying
\[
    \forall B\in\mathcal X_0:\quad
    \mathrm{Safe}^{L'}_T(B,\overline\pi).
\]
Furthermore, suppose that for every box $B$ encountered
in its unmerged continuation with remaining
horizon $h$, the backend explores
$\overline\pi(q(B))\in\mathcal A_{\mathcal W}(B,h)$.
Then \method returns a certified safe controller.
\end{restatable}

Theorems~\ref{thm:soundness}, Theorem~\ref{thm:termination} and Corollary~\ref {cor:backend-completeness} are directly applicable to our ACC and Car2D experiments. Docking2D employs an additional greedy restart optimization; however, with minor modifications, all theorems extend to this setting. In particular, one needs to account for an additional outer loop that only returns once a repair-free pass has been performed. Hence, the theoretical guarantees also hold for its experiments. 

\method attempts to synthesize a controller that remains close to the original
policy $\pi_0$ in terms of the $L_1$-distance in the popcount integer space. However, finding
an optimal repair is generally computationally hard: even minimal strategy
repair in finite reachability games is NP-complete~\citep{gaillard2023strategy}.
Hence, \method employs heuristics and accepts non-optimal assignments after a
search timeout.

\section{Experiments}
\makeatletter
\newcommand{\wrapfigcleanup}{\WF@finale}
\makeatother
\begin{wrapfigure}[34]{r}{0.4\textwidth}
\centering
\vspace{-2\baselineskip}
\includegraphics[width=\linewidth]{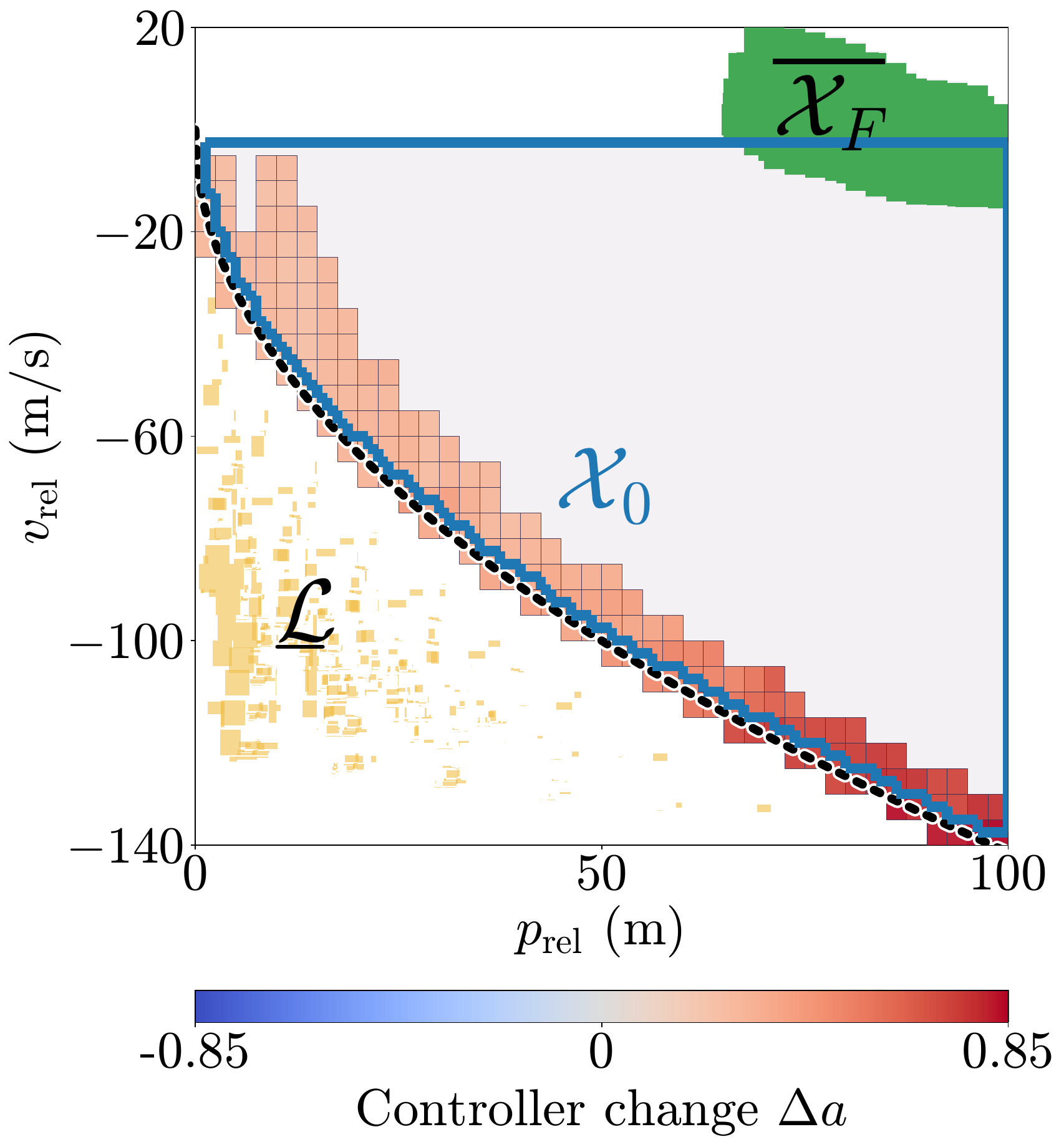}
\vspace*{-2\baselineskip}
\caption{ACC ($\mathcal P_{max}=100$, with $\mathcal W_{brake}$): Showing $\mathcal X_0$,  the repaired cells with the associated magnitude change, and the final fixed-point $\overline{\mathcal{X}}_F$. The learned losing set $\underline{\mathcal L}$ is shown in orange. Changes in the controller are localized close to the losing set boundary, which \method was not provided with in this run.} 
\label{fig:reachacc}
\vspace{0.2\baselineskip}
{
\captionof{table}{ACC: verification + repair runtime~[s] at $ \mathcal P_{max}=30$ over 5 seeds.}
\label{tab:acc-runtime-x30}
\begin{tabular}{lr}
\toprule
Method & t [s] \\
\midrule
No losing-set               & $163 \pm 103$ \\
Online losing-set           & $29 \pm 20$ \\
Initial losing-set          & $11 \pm 5$ \\
$\mathcal W_{brake}$  & $ 5 \pm 1$ \\
$\mathcal L^*$                   & $3 \pm 1$ \\
\bottomrule
\end{tabular}}
\end{wrapfigure}
We investigate three tasks, showing different properties of \method across them: \emph{Adaptive Cruise Control (ACC)}~\citep{fulton2018safe,teuber2024provably}, \emph{Docking2D} from the AINNCS competition~\citep{lopez2025archcomp25ainncs}, and \emph{Car2D}, which uses the NAV dynamics from~\citep{lopez2025archcomp25ainncs}. We briefly summarize Docking2D and Car2D here and defer their full analysis to Appendix~\ref{sec:additional}. All experiments were performed on an AMD EPYC 9534 (Zen 4) CPU, with an NVIDIA L40S GPU used where applicable. In particular, \method's reachability analysis runs on the GPU. Our implementation is available on
\href{https://github.com/FKresse/reacheco}{GitHub}.

\subsection{Adaptive Cruise Control}

\paragraph{Setup.} A controlled vehicle must avoid a constant-velocity lead vehicle while maintaining
comfortable acceleration. Relative distance and velocity are denoted by
$p_{\mathrm{rel}}$ and $v_{\mathrm{rel}}$, respectively, with
$v_{\mathrm{rel}}<0$ indicating that the vehicles approach each other. We train five DWCs following \citet{kresse2026differentiable}.
The system has a known losing set $\mathcal L^*$ (Figure~\ref{fig:reachacc}, everything below the dotted black line): states above its
boundary can avoid collision by braking, whereas in states below the line, an eventual collision is unavoidable. We initialize \method with boxes entirely outside 
$\mathcal L^*$ satisfying
$v_{\mathrm{rel}}<0$ and $p_{\mathrm{rel}}<\mathcal P_{max}$, covering $\approx96\%$ of the corresponding safe region. The unsafe set $\mathcal U$ is defined as $p_{rel}\leq 0$, corresponding to a rear-end collision between the lead and controlled vehicle. All \method runs use $T=200$, and all runs reach a
closed-loop fixed point within this budget, certifying infinite-horizon
safety.  Details of the dynamics and training are given in Appendix~\ref{app:environments}.
\paragraph{Losing set knowledge reduces runtime.}
We cumulatively evaluate four levels of losing set knowledge:
(i) no $\underline{\mathcal{L}}$ expansion, (ii) $\underline{\mathcal{L}}$ expansion, 
(iii) initialization with an offline-computed
 $\underline{\mathcal{L}}$ fixed point, and (iv) the exact braking invariant.
$\mathcal W$ uses 11 discrete actions including
zero and both extrema. We additionally evaluate a single-action full-braking backend $\mathcal W_{brake}$, under (ii). 

Table~\ref{tab:acc-runtime-x30} shows that stronger safety information reduces
repair runtime. For $\mathcal P_{max}=30$, runtime
drops from $163$s without losing-set learning to $29$s with online learning,
$11$s with an underapproximated initialized losing set, and $3$s with $\mathcal L^*$.
The full-braking and invariant-based variants additionally scale to $\mathcal P_{max}=100$~(Table \ref{tab:acc-runtime}). At this initial set size, games of depth-11+ are necessary, making $\mathcal W$ infeasible. As shown in Figure~\ref{fig:game_depths} in the Appendix: losing-set
expansion eliminates a large number of deep continuation games. At
$\mathcal P_{max}=30$, this reduces depth-3--7 games by a large amount, while the knowledge of $\mathcal L^*$ requires only depth-1 games.
\paragraph{Repairs are localized.}\method successfully repairs all five DWC seeds and increases their mean
reward from $2434$ to $3861$; per-seed results are reported in
Table~\ref{tab:acc-evaluation} in the Appendix. Qualitatively, Figure~\ref{fig:reachacc} shows
that the resulting modifications concentrate near the boundary of
$\mathcal L^*$, even when \method is provided only with the unsafe set
$\mathcal U$. Table~\ref{tab:acc-runtime} further shows that access to
$\mathcal L^*$ and the more general backend $\mathcal W$ reduce both the number 
of modified policy cells and the mean
action change. 
\paragraph{Other benchmarked methods require $\mathcal L^*$ or do not certify.}
We compare against methods with and without access to $\mathcal L^*$.
When $\mathcal L^*$ is available, we construct counterexample-guided repair
loops using either BBReach~\citep{tian2023taming} for DWCs or
NCubeV~\citep{teuber2024provably} for ReLU controllers. Verification
identifies failing regions, based on which the controller is repaired and
reverified until the first successful certificate is obtained.
For supervised repair, we fine-tune on the identified failures together with
safe transitions from the original controller to reduce unintended changes; 
full details are given in Appendix~\ref{app:environments}. 
For the ReLU controller, we additionally combine NCubeV with
REASSURE~\citep{fu2022reassure}, which constructs a local patch network from
the identified unsafe regions. We cannot certify the resulting patched
network with NCubeV because its architecture is outside the verifier's
supported network class, and therefore report the resulting repair size
without a closed-loop certificate. Among the methods that obtain a certificate of safety, \method is faster 
and produces smaller behavioral changes than the supervised fine-tuning 
baselines, requiring only a few hundred additional LUT6s to repair the DWC.
REASSURE style patches yield a smaller amount of changes. This is at least in part due to the fact that the underlying ReLU network is already safer than the initial DWCs.  Nevertheless, it requires a patch size of 2350 additional \emph{floating-point} parameters. 

In a second setup, we assume that only the unsafe set is known; for this setting, we use STLGym~\citep{hamilton2022training} to fine-tune the original controller for an additional $200{,}000$ time steps. Figure~\ref{fig:acc-stlgym-training} in the Appendix shows the resulting reward and empirical safety across training. Training is unstable across seeds: in some runs, reward degrades substantially, and none of the resulting controllers achieves $100\%$ empirical safety.

\begin{table}[t]
\centering
\caption{ACC at $\mathcal P_{max}=100$. $t$ denotes the runtime for verification plus repair.
If verification fails, $t$ reports repair time only. Runtime and repair size are
reported over five trained model seeds. $^\dagger$ denotes a ReLU network.
For \method, \emph{Param.} denotes the number of patch LUT6s; for NCubeV+REASSURE,
results are reported only for the network from \citep{teuber2024provably}, and
\emph{Param.} denotes the number of additional floating-point parameters. The first column indicates the losing set information provided: $\mathcal L^*$ denotes the known losing set, $\mathcal U$ denotes only the unsafe set.}
\label{tab:acc-runtime}
\begin{tabular}{clccccc}
\toprule
& & & \multicolumn{3}{c}{Repair} \\
\cmidrule(lr){4-6}
& Method & Verified & $t$ [s] & Param. & Change [\%] & mean $|\Delta|$ \\
\midrule

\multirow{4}{*}{\rotatebox[origin=c]{90}{$ \mathcal{L}^*$}}
& \method
& $\checkmark$
& $2_{\pm 1}$
& $129_{\pm 24}$
& $8.2_{\pm 2}$
& $0.036_{\pm 0.014}$ \\

& BBReach + Supervised
& $\checkmark$
& $25_{\pm 2}$
& $0$
& $96.9_{\pm 7.0}$
& $0.324_{\pm 0.119}$ \\

& NCubeV + Supervised$^\dagger$
& $\checkmark$
& $1760_{\pm 1188}$
& $0$
& $47.77_{\pm 4.51}$
& $0.036_{\pm 0.012}$ \\

& NCubeV + REASSURE$^\dagger$
& $?$
& $88$
& $2350$
& $1.2$
& $0.003$ \\

\midrule

\multirow{2}{*}{\rotatebox[origin=c]{90}{$\mathcal{U}$}}
& \method ($\mathcal W_{\mathrm{brake}}$)
& $\checkmark$
& $19_{\pm 3}$
& $211_{\pm 18}$
& $12.6_{\pm 3.2}$
& $0.057_{\pm 0.023}$ \\

& STLGym$^\dagger$
& \xmark
& $47_{\pm 2}$
& $0.0$
& $41.4_{\pm 7.2}$
& $0.151_{\pm 0.043}$ \\

\bottomrule
\vspace{-0.7cm}
\end{tabular}
\end{table}
\subsection{Docking2D and Car2D: Takeaways}
The main analysis of these two case studies is deferred to Appendix~\ref{sec:additional}, we highlight main results.
\paragraph{Docking2D: Reachability enables \emph{highly} localized patches.}
Docking2D has a large continuous state space and long closed-loop horizons.
Nevertheless, \method certifies all our three DWC controllers, till docked, while modifying only a small 
number of cells. For the seed exhibiting the most empirically unsafe transitions, only 3,309 policy
cells are changed, corresponding to less than $10^{-3}\%$ of the overall policy cells.

\paragraph{Car2D: enabling certification through small changes.}
The original controller exhibits no unsafe behavior in 1,000 sampled rollouts,
yet closed-loop reachability fails to certify it, identifying a potential safety
violation after three timesteps. \method repairs the controller in 42\,s,
requiring continuation games of up to depth~3 while modifying only 39 policy
cells. Several repaired regions are almost never visited by the sampled
trajectories. These regions may correspond either to rare behaviors missed by
sampling or to conservatism introduced by the reachability overapproximation;
in either case, modifying them is sufficient to make the controller certifiable.

\section{Summary and Discussion}

We introduced \method, which enables targeted repair of abstraction-based neural network controllers
to fulfill formally certified safety conditions. 
On a technical level, \method brings the constraint-based view established in 
static neural-network repair to the closed-loop setting by explicitly accounting
for the future behavior induced by controller modifications. Repair constraints
are derived lazily from failed forward reachability analysis and subsequent
finite-horizon continuation reasoning. This formulation enables strong formal
guarantees on the repair procedure, including termination and, for an exhaustive
variant, completeness relative to the reachability abstraction. 
For the special class of Differentiable Weightless Controllers (DWCs), we further 
showed that the resulting repairs can be compiled into exact circuit patches. 
Because \method only modifies regions encountered during closed-loop reachability 
analysis, the resulting repairs can remain highly localized in the controller's 
input space. At the circuit level, the additional logic consists primarily of 
LUT6s, allowing the repair to be realized as a lightweight overlay on the original 
controller.

Some limitations remain: \method relies on \emph{deep continuation games}, 
which can become computationally expensive for long-horizon problems.
This can be mitigated by additional prior information, such as knowledge of 
the losing region. Our experiments show that even partial losing-set information 
can substantially reduce the required search. 
Even though repair patches are highly localized in the controller's input space,
many of them might be needed for originally unsafe controllers. A promising direction
for future work is to exploit additional structure during patch compilation. 

Our current implementation additionally relies on relatively coarse, 
benchmark-specific post-image and inter-sample tube computations chosen for
efficient GPU execution. 
Tighter GPU-based reachability procedures could reduce conservatism, while 
more general tube constructions would make the approach easier to transfer 
across dynamical systems. 
Finally, our experiments consider relatively low-dimensional systems, as is 
common in closed-loop reachability benchmarks. Scaling both verification 
and the additional repair search to higher-dimensional dynamics remains the
most important future challenge. 

\newpage

\subsection*{AI use statement}

In this work, we used generative AI tools for developing conceptual frameworks, formulating mathematical claims, assisting in the writing of the presented proofs, for providing feedback on research methodology and experiments, implement methods, and interpreting results.
We have not used generative AI tools for proposing or refining hypotheses, supporting qualitative and thematic data analysis, or providing critical ingredients for proving mathematical claims. Synthetic dataset generation, assisting with translation, cleaning and reformatting datasets are not applicable to this work.

Additionally, we used generative AI tools to assist with figure generation, code prototyping, drafting and polishing text, brainstorming, searching for information, and proposing keywords for the paper. The tube overapproximations used in this work were initially derived with assistance from generative AI tools. We reviewed all AI-assisted work. The original prototype implementation was extensively audited, rewritten, and restructured, and AI-provided technical information was carefully reviewed and independently verified by at least one of the authors. We also conducted conventional literature searches to verify and complement AI-assisted information retrieval. We take responsibility for the final content of this work, including text, claims, code, and other artifacts produced with the aid of generative AI.

\subsection*{Reproducibility statement}
Our implementation is available on
\href{https://github.com/FKresse/reacheco}{GitHub}.
Implementation details are provided in
Appendix~\ref{sec:formal}.
We have provided proofs of our Theorems in Appendix~\ref{app:proofs}.
Furthermore, we have provided detailed descriptions of all case study 
environments we investigate in Appendix~\ref{app:environments}.

\bibliography{iclr2027_conference}
\bibliographystyle{iclr2027_conference}

\appendix

\clearpage
\section{Additional Experiments}
\label{sec:additional}
\subsection{Docking2D}
\label{sec:docking2d}
\begin{table}
    \caption{Docking2D:  $\mathcal X$ reports empirical safety for 1000 rollouts from the 130 meter starting circle. $\mathcal X_0$ reports empirical safety from the initial region (before repair). $t$ reports the overall time for both repair and verification. tanh denotes the controller utilized in~\citep{lopez2025archcomp25ainncs}. Param reports the increase in size in terms of LUT6s of the repaired controller. Note that naive repair is an estimated lower bound and not synthesized.}
    \label{tab:docking2d}
\centering
\begin{tabular}{ccccccrrr}
\toprule
& \multicolumn{2}{c}{Emp.\ safe [\%]} & \multicolumn{2}{c}{Reward} & & & \multicolumn{2}{c}{Param [LUT6]} \\
\cmidrule(lr){2-3}\cmidrule(lr){4-5}\cmidrule(lr){8-9}
$\pi$ & $\mathcal{X}$ & $\mathcal{X}_0$ & original & after repair & Verified & t [h] & ours & naive \\
\midrule
tanh & 95.2 & 100.0 & $3.096 \pm 0.069$ & $\cdot$ & \xmark & $\cdot$ & $\cdot$ & $\cdot$ \\
\midrule
1 & 100.0 & 100.0 & $3.066 \pm 0.067$ & $3.066 \pm 0.067$ & \cmark & 0.25 & +50 & +92k \\
2 & 95.6 & 100.0 & $2.994 \pm 0.061$ & $2.994 \pm 0.061$ & \cmark & 0.38 & +0 & +87k \\
3 & 96.0 & 75.9 & $3.003 \pm 0.148$ & $3.060 \pm 0.066$ & \cmark & 1.09 & +6.4k & +120k \\
\bottomrule
\end{tabular}
\end{table}\paragraph{Setup.}
Docking2D is a challenging benchmark for neural-network safety controller verification. For example, none 
of the entries to the ARCH-COMP AINNCS competition in 2025 was able to certify the full provided initial region~\citep{lopez2025archcomp25ainncs}. We consider a larger initial region, covering the competition's region, and, moreover, verify the reach-avoid property proposed in~\citep{mandal2024formally}, rather than only the original 40-step ARCH-COMP specification. The resulting set propagation requires more than \emph{850 time steps} before all sets have converged to the goal region around the origin.

We train three DWC policies with PPO; training details are given in Appendix~\ref{app:environments}. For continuation search, $\mathcal W$ uses a backup controller that targets the docking point $(0,0)$ at a desired speed of $0.15\mathrm{m/s}$ and proposes the corresponding discrete action. The trained DWCs achieve reward close to the competition tanh policy, with some seeds remaining empirically unsafe before repair. We split the initial set into four smaller initial boxes. To reduce repeated recomputation, we use a greedy restart strategy: after a repair, verification resumes from the depth at which the corresponding continuation game was constructed rather than from the earliest potentially affected time step. Since certification is reported only after a subsequent repair-free pass over all initial boxes, this optimization does not weaken the final guarantee. Our main insights are:

\begin{wrapfigure}[18]{r}{0.4\textwidth}
    \centering
    \includegraphics[width=\linewidth]{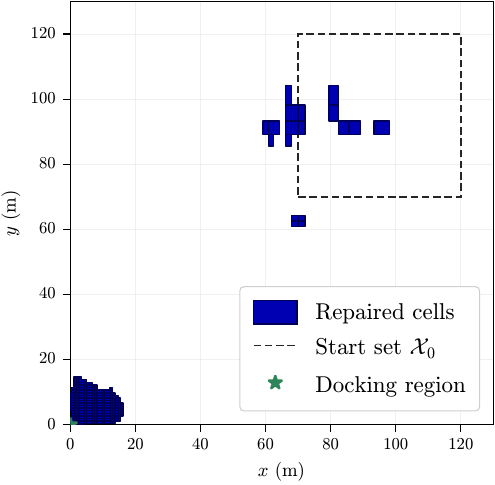}
    \vspace{-0.7cm}
    \caption{Docking2D: repaired regions of the DWC seed 3 projected onto the $(x,y)$ plane.}
    \label{fig:docking2d-repaired-regions}
\end{wrapfigure}
\paragraph{Depth-1 games are insufficient.} When performing repair, we observe a total of $21$ depth-2 games. These occur when the intersample tube does not allow a safe continuation, requiring backtracking.

\paragraph{Repairs are \emph{highly} localized.}
Table \ref{tab:docking2d} reports reward, repair, and verification time as well as additional LUT6s
for patch circuits. Seed~1 is the only controller that is empirically safe over the entire
evaluation region, yet still requires modifications to $5$ policy cells before
it can be certified. Seed~2 exhibits empirical failures outside the certified
initial region but none from $\mathcal X_0$; \method can certify 
it without having to modify the controller at all. Seed~3 is empirically unsafe and
requires the largest circuit patch, adding roughly $6.4$k LUT6s. Nevertheless,
only $3{,}309$ policy cells are modified, corresponding to less than
$10^{-3}\%$ of the overall cells. Thus, even when the compiled
circuit patch is comparatively large, the behavioral modification remains
extremely local. Figure~\ref{fig:docking2d-repaired-regions} visualizes these
changes and shows that they are concentrated primarily around the docking region. 

\paragraph{Reachability information \emph{substantially} reduces patch size.}
We compare against a naive patching scheme that does not use
reachability-derived information, but instead patches every fully safe policy
cell whose transition can enter the unsafe set from anywhere in the safe
region. As shown in Table~\ref{tab:docking2d}, the resulting patches are
substantially larger than those produced by \method, requiring up to several
orders of magnitude more LUT6s. Moreover, these numbers are conservative:
the naive construction considers only cells fully contained in the safe set
and therefore underapproximates the patch that would be required over the
entire boundary.

\subsection{Car2D}

\paragraph{Setup.}
\begin{wrapfigure}[18]{r}{0.4\textwidth}
\centering
\includegraphics[width=\linewidth]{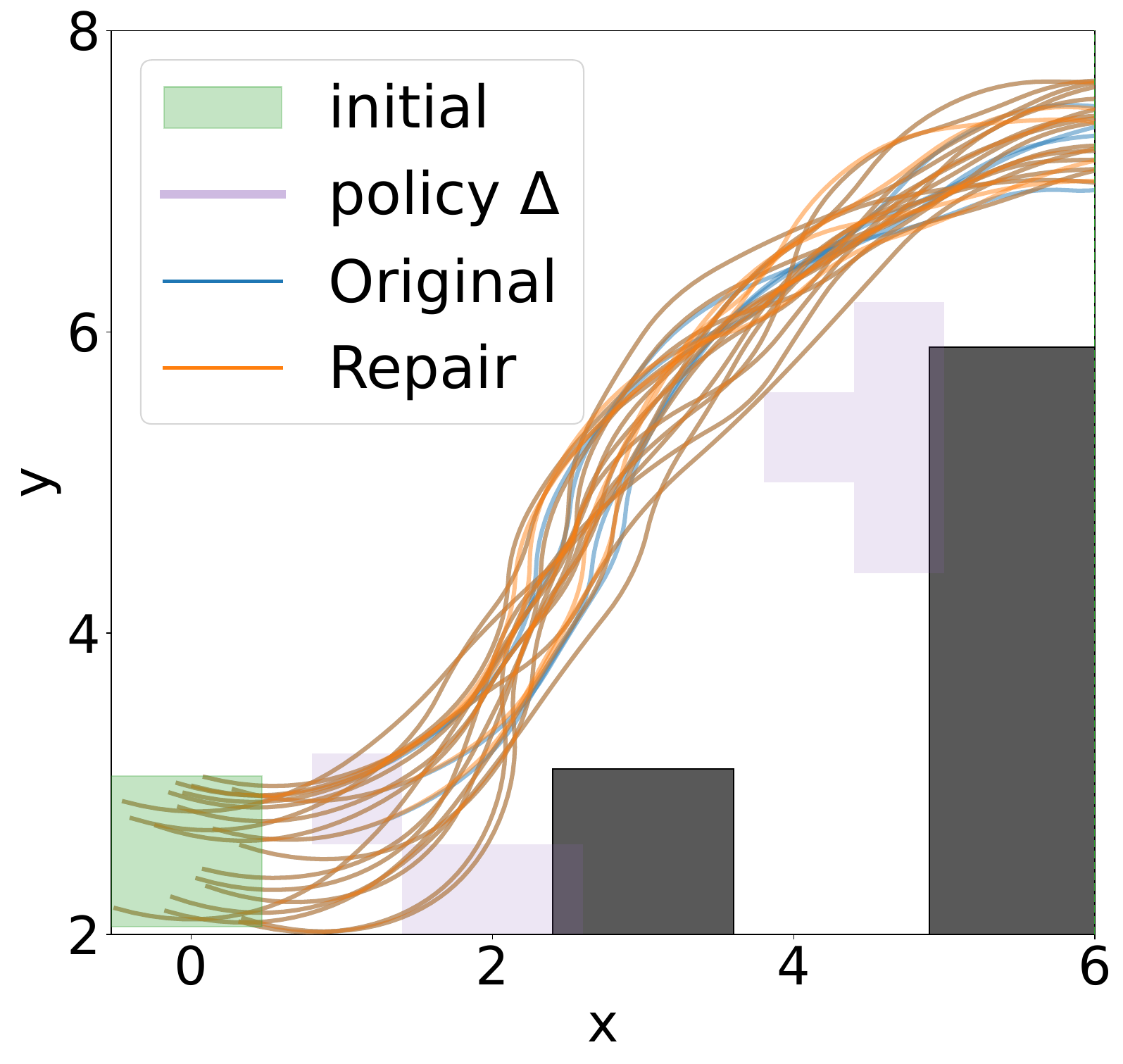}
\vspace{-0.8cm}
\caption{Car2D rollouts before and after repair, with modified policy cells highlighted.}
\label{fig:car2d-repaired}
\end{wrapfigure}

We investigate a modified version of the nonlinear
NAV benchmark from the ARCH-COMP AINNCS competition~\citep{lopez2025archcomp25ainncs}.
We train a DWC controller with TD3~\citep{fujimoto2018addressing}, with specifics given in Appendix~\ref{app:environments}. The
controller must reach the
goal region $x>6$ while avoiding the two obstacles. Figure~\ref{fig:car2d-repaired} shows the
initial set and the obstacles. During repair, we modify only the
steering output and consider a repair alphabet consisting of three
increasingly large steering corrections in each direction, together with
the two extreme steering actions.

\paragraph{Repair restores certifiability.}
The original controller is empirically safe over $1000$ rollouts sampled
from the initial set, exhibiting no observed unsafe
trajectories. Nevertheless, formal verification with our BBReach implementation finds a
reachable-set intersection with the unsafe set at time step $t=3$.
We subsequently apply \method. During repair, \method constructs
$1358$ depth-$1$, $88$ depth-$2$, and $7$ depth-$3$ continuation games.
After $42\,\mathrm{s}$, \method returns a repaired controller modifying $39$ policy cells that passes verification. Figure~\ref{fig:car2d-repaired} compares rollouts of the original and
repaired controllers and visualizes the modified policy cells. Notably, the upper repaired regions are virtually absent from the sampled
rollouts. Their modification is nevertheless required for the reachability
analysis to pass.
\section{Appendix: Related Work}
\label{app:related work}

\paragraph{Shielding.} Shielding enforces safety at runtime by monitoring the nominal controller
and overriding actions when necessary~\citep{bloem2015shield,alshiekh2018safe}.
Shields may be synthesized offline or perform additional safety reasoning
online~\citep{bastani2021safe,konighofer2022online}.
\citet{corsi2024verification} use offline verification to identify regions
in which an online shield needs to be active.
More recently, \citet{maderbacher2025adaptive} propose an architecture in 
which a performance-optimized controller is permitted to act only inside
a certified recoverable region, while a baseline controller intervenes
otherwise.

Our repaired controller is closely related to a sparse, statically compiled
shield: it overrides the nominal policy only on a finite set of state regions.
The main distinction is that \method uses failed closed-loop reachability
verification to discover the required interventions offline, incorporates
them directly into the nominal controller, and reverifies the resulting
controller. Consequently, deployment requires neither runtime safety
reasoning nor invocation of a separate backup controller.

\paragraph{Closed-Loop Verification of Neural Network Controllers}

Safety verification of neural-network controlled systems is often performed with reachability analysis: an over-approximation of the set of reachable closed-loop states is propagated forward in time and checked for intersection with an unsafe set. A large ecosystem of methods and tools implements this basic recipe with different set representations and abstractions of the network~\citep{xiang2018reachability,ivanov2019verisig,ivanov2021verisig,dutta2019sherlock,tran2020nnv,lopez2023nnv,huang2022polar,hu2020reach,schilling2022verification,sidrane2022overt,tian2023boosting,tian2023taming}. The ARCH-COMP AINNCS category provides a yearly comparison between some of these tools~\citep{lopez2025archcomp25ainncs}. A second line of research focuses on certificate-based verification, reducing the need to argue about entire trajectories to single-step queries on the certificate~\citep{chang2019neural,zhao2020synthesizing, mandal2024formally,edwards2024fossil}. However, this adds the need to both synthesize an appropriate certificate and subsequently check it. 

Another reachability-related line of work trains controllers that are constant on a finite partition of the state space, so that closed-loop verification never has to perform over-approximation of the neural network. \citet{jin2022trainify} introduce Trainify, where a neural network controller is trained in a CEGAR loop that iteratively refines an abstraction of the state space until the control objectives can be fulfilled; since actions are constant per abstract state, neural network verification is skipped entirely and model checking is performed. \citet{tian2023taming} instead fix an initial abstraction, train on it, and perform reachability analysis (BBReach), again exploiting that constant actions make over-approximating the network unnecessary. \citet{tian2023boosting} extend the controller class to piecewise-affine decision networks, enabling coarser abstractions, and \citet{yang2025formal} extend abstraction-based controller verification to probabilistic safety.

\paragraph{Constraint based Repair.} Counterexample-guided finetuning is not guaranteed to terminate \citep{boetius2023robust}. Hence, there has been work to provide stronger repair guarantees by modifying the internal parameters of a network based on input-output constraints~\citep{goldberger2020minimal, sotoudeh2021provable,  bauer2022specrepair,tao2023architecture, sun2025autoric} or adding patches to it~\citep{fu2022reassure}. These approaches reason about the network specification itself rather than
the closed-loop system dynamics.
\paragraph{Local Controller Repair.}
Encompasses methods that rely on \emph{single} step reasoning, either performing repair based on counterexamples~\citep{santa2022safe} or one-step reachability sets~\citep{yang2022neural}. \citet{teuber2024provably} utilize a control envelope to transform the global controller safety problem into a local one. However, this requires the envelope to be available. \citet{santa2022safe}, perform constraint optimization at a known counterexample while safeguarding certain global properties. \citet{mandal2024formally} reduce the multistep repair problem to a single-step one by fine-tuning a Lyapunov-barrier certificate together with the controller; this shifts the computational complexity to checking the certificate.

\paragraph{Trajectory-Level Controller Repair.} Many global controller repair methods rely on parameter fine-tuning, often based on STL robustness~\citep{maler2004monitoring, fainekos2009robustness}. Examples are STLGym~\citep{hamilton2022training} and ISAR~\citep{lu2025repairing}. Other methods rely on discovering failing trajectories and supervised fine-tuning ~\citep{lyu2023autorepair} or direct changes to the policy discovered through exhaustive reasoning~\citep{lu2023causal}.
Lastly,~\citet{zhou2020runtime} and \citet{yu2025neural} utilize a runtime monitor and perform fine-tuning of the controller based on runtime feedback.

\paragraph{Set-based training}
A set of methods attempts to improve the likelihood that a set-based closed-loop proof will succeed~\citep{wang2021verificationloopcorrectbyconstructioncontrol,vlantis2023failing,chaudhury2025learning,shen2026parallel}. In these methods set-based information is included in the training loss, aiming to reduce over-approximation during verification.

\paragraph{Circuit Rectification}
Repairing a deployed circuit with a small patch instead of re-synthesizing it is long-standing practice in logic design, where engineering change orders (ECOs) rectify an implementation towards a revised specification with minimal modification~\citep{lin1995logic,krishnaswamy2009deltasyn}. Our patch compilation adopts this philosophy for learned controllers, with one key difference: the revised specification is not written by a designer but generated automatically, in the form of action-correction constraints discovered by a closed-loop reachability verifier.

\paragraph{Strategy Repair}

Synthesis based on discrete-reachability games has a long history~\citep{de2007concurrent}. However, even for discrete state-spaces, prior work on repair has addressed the problem of \emph{strategy repair} only relatively recently \citep{gaillard2023strategy,calamoneri2026strategy}. \method addresses an analogous problem for learned controllers that operate on continuous dynamics, however, in our setting the underlying game is not available a priori: relevant transitions are constructed lazily from failed closed-loop reachability verification.

\section{Formal Definitions and Algorithm Details}
\label{sec:formal}
\subsection{Reachability}
As in the main body of text: we study a task with a  continuous state space $\mathcal{X}\subseteq\R^d$ and system dynamics governed by a 
differential equation, $\dot x = f(x,a)$, where $a\in\mathcal{A}$ 
is a control action. 

Now, let $\Phi_t^a(s)$ denote the state reached after time $t$ from $s$ 
under constant input $a$.
For a set of states, $S\subseteq\mathcal X$, and an action $a\in\mathcal A$, define 
\begin{equation}
\Post(S,a)
:=
\{\Phi_\Delta^a(x)\mid x\in S\},
\qquad
\Tube(S,a)
:=
\{\Phi_s^a(x)\mid x\in S,\ s\in[0,\Delta]\}.
\end{equation}

Because of the continuous nature of the underlying dynamics, exact reachability cannot be computed in general. We therefore rely on \emph{sound over-approximations}. 
$$
\overlinePost(S,a)\supseteq\Post(S,a),
\qquad
\overlineTube(S,a)\supseteq\Tube(S,a).
$$
\subsection{Problem Setting}
We now formally define \emph{piecewise-constant abstraction-based controller}, as they are the controller type studied throughout.
\begin{definition}[Piecewise-constant abstraction-based controller]
\label{def:pcabc}
A controller, $\pi:\mathcal X\to\mathcal A$, is called \emph{piecewise-constant 
abstraction-based}, if $\mathcal A\subseteq\mathbb R^m$ is finite and 
exposes a finite axis-aligned policy cell partition $\mathcal Q$ of $\mathcal X$ 
such that for any $q \in \mathcal Q:\  x,x'\in q \Rightarrow \pi(x)=\pi(x')$.
Hence, $\pi$ equivalently induces a finite map $\pi:\mathcal Q\to\mathcal A$,
and we write $\mathcal A^{\mathcal Q} := \{\pi \mid \pi:\mathcal Q\to\mathcal A\}$ for the set of all controllers over the fixed policy partition $\mathcal Q$
and action alphabet $\mathcal A$. 
\end{definition}

Let us now formally define the repair problem.

\begin{definition}[Repair instance]
\label{def:repair-instance}
A \emph{repair instance} consists of an abstraction-based controller
$\pi_0$, a finite set of continuation actions
$\mathcal A_{\cont}\subseteq\mathcal A$, sound propagation operators
$\overlinePost$ and $\overlineTube$, an initial set
$\mathcal X_0\subseteq\mathcal X$, given as a collection of boxes, an unsafe set
$\mathcal U\subseteq\mathcal X$, an optional absorbing goal set
$\mathcal {X_G}\subseteq\mathcal X$, and a verification budget $T$.
\end{definition}

\begin{definition}[Repair class]
\label{def:repair-class}
Given a repair instance, set
$\mathcal A(q)=\{\pi_0(q)\}\cup\mathcal A_{\cont}$ for every
$q\in\mathcal Q$. Then
\[
    \Pi_R:=\prod_{q\in\mathcal Q}\mathcal A(q)
\]
is the \emph{repair class} of $\pi_0$.
It contains all controllers that either retain the original action or select
a continuation action independently on each policy cell.
\end{definition}

\begin{definition}[Certified repair]
\label{def:certified-repair}
A \emph{certified repair} of a repair instance is a controller
$\pi'\in\Pi_R$ that is certified safe according to
Definition~\ref{def:certified-safety}.
\end{definition}

\subsection{Controllability and Losing Regions}
\label{sec:controllable-envelopes}

Given a failing reachable region, repair must determine how many time steps before the reported violation an intervention is needed. Whether an intervention at a
given state can still prevent the violation depends on whether safety can be
maintained from the resulting successor states. For the infinite time case this is captured by the notion of \emph{controlled invariance}
\citep{bertsekas1972infinite,blanchini1999set}; its maximal set corresponds to
the \emph{viability kernel} in viability theory
\citep{aubin1990survey} and to the \emph{winning region} of a safety game
\citep{tabuada2009verification}.

We define controllability relative to the finite continuation alphabet
$\mathcal A_{\cont}$ used by repair.

\begin{definition}[Controlled invariant set]
A set $C\subseteq \mathcal S$, with
$\mathcal S:=\mathcal X\setminus\mathcal U$, is \emph{controlled invariant} if
\[
    \forall x\in C\;\exists a\in\mathcal A_{\cont}:
    \qquad
    \Tube(\{x\},a)\cap\mathcal U=\emptyset
    \;\land\;
    \Post(\{x\},a)\subseteq C.
\]
The maximal controlled invariant subset of $\mathcal S$ is denoted
$\mathcal C^*$.

\end{definition}
Dually, let \(
    \mathcal L^* := \mathcal X \setminus \mathcal C^*
\) denote the \emph{losing set}, i.e., the states from which safety cannot be
maintained indefinitely using only actions from $\mathcal A_{\cont}$.
If a controlled invariant set is known, it yields a one-step condition for repair: the repaired action must keep the current transition
safe and map all successors into the invariant.

When $\mathcal C^*$ is unavailable, the one-step repair condition above
cannot be applied directly. \method instead starts from a sound
under-approximation
$\underline{\mathcal L}\subseteq\mathcal L^*$, which may initially contain
only the unsafe set $\mathcal U$. 

Hence, we define the \textsc{SafeStep} predicate in eq. \ref{eq:safe step} as the verifiers analogue to the ideal one-step condition above. Since $\underline{\mathcal L}$ may be a strict under-approximation of
$\mathcal L^*$, however, satisfying $\mathrm{SafeStep}$ may still lead to a safety violation in later time steps.

 For later use, let us now additionally define the predicate $\mathrm{Safe}_h$.
 For a box $B$ contained in policy cell $q(B)$ and a fixed controller $\pi$, 
$\mathrm{Safe}_0(B,\pi):=\mathrm{true}$ and, for $h\ge1$, define
\begin{equation}
\label{eq:safe-h}
\begin{aligned}
    \mathrm{Safe}_h(B, \pi)
    \;:\Longleftrightarrow\;&\mathrm{SafeStep}(B,\pi(q(B)))\\
    &\land\;\forall B'\in\Succ(B,\pi(q(B))):
    \mathrm{Safe}_{h-1}(B',\pi).
\end{aligned}
\end{equation}
Thus $\mathrm{Safe}_h$ checks the continuation under $\pi$ itself. To make the dependence on the losing set in $\mathrm{SafeStep}$ explicit, we write
$\mathrm{Safe}^{L'}_h(B,\pi)$ evaluated with $\underline{\mathcal L}=L'$ in $\mathrm{SafeStep}$.

\subsection{Algorithm Details}
\paragraph{Forward Verification.}
ReachECO uses our implementation of BBReach for forward reachability analysis~\citep{tian2023taming}, 
to propagate the current repaired
controller $\pi'$ in \textsc{ReachStep}. At time step $k$, 
the initial failing frontier, the set of \emph{obligations}, is 
\begin{equation}
\label{eq:frontier}
    \mathcal F
    :=
    \left\{
        (B,1)
        \;\middle|\;
        B\in \overline{\mathcal{X}}_k\;\land\;
        \neg\mathrm{SafeStep}(B,\pi'(q(B)))
    \right\}.
\end{equation}
A nonempty frontier triggers \textsc{BRepair}. After a repair or an expansion of
$\underline{\mathcal L}$, verification resumes from the earliest affected
time step. If a reachable-set fixed point is reached before the horizon budget $T$, we obtain an unbounded safety certificate. Furthermore, intersections with the absorbing goal are removed from further reachability steps.

\paragraph{Notes on Merging.}
Merging boxes that belong to the same policy cell is essential for the
scalability of BBReach-based reachability analysis~\citep{tian2023taming}.
Without merging, every propagated box has a unique parent in the preceding
timestep. Merging destroys this property: a merged box may contain successors
originating from several distinct parent boxes.

To retain the efficiency of merging while recovering an exact lineage when a
safety violation is encountered, \method records the candidate parents contributing to
each merged box. Upon detecting a failing merged box, it replays only these
candidate parents from the preceding timestep, with merging disabled. Thus,
this refinement is performed lazily and only along the portion of the
reachability computation relevant to the safety violation. If no safety violation remains after
replay, the original violation was induced by merging. The corresponding
obligation is removed from the failing frontier, and a \emph{merge mask} is
updated with the timestep and policy-cell identifier at which merging must be
disabled in subsequent forward propagations.

Otherwise, replay yields one or more genuinely failing unmerged boxes, each
with a unique parent. Consequently, \textsc{Backtrack} is only applied to
unmerged boxes: for an obligation $(B,h)$, if $B_{\mathrm{pred}}$ is the unique
parent of $B$ in the preceding timestep, \textsc{Backtrack} replaces
$(B,h)$ by $(B_{\mathrm{pred}},h+1)$. Repeating this operation therefore
recovers a unique predecessor lineage despite the use of merging during the
original forward propagation.

\paragraph{Repair Constraints.}

Recall a \emph{repair constraint}
\(
    \gamma=(\mathcal G,i),
\)
with root timestep $i$ and game DAG $\mathcal G$. 
Let $\mathcal A_v$ denote the set of actions stored at node $v$,
which allow a safe continuation.
For a controller $\pi$, the constraint is satisfied recursively as follows:
at a node $v$, the action $a=\pi(q(B_v)) \in \mathcal A_v$, and, if $h_v>1$, the 
same condition must hold for every child node
corresponding to a box
$B'\in\Succ(B_v,a)$.
We write
\[
    \pi\models\gamma
\]
if this condition holds from the root of $\mathcal G$.

For a set of constraints $\Gamma$, we write
\begin{equation}
    \pi\models\Gamma
    \quad\Longleftrightarrow\quad
    \forall\gamma\in\Gamma:\ \pi\models\gamma .
\end{equation}

Growing $\underline{\mathcal L}$ does \emph{not} reinterpret
existing games. Additional exclusions are performed by forward
reverification and newly generated constraints.

\paragraph{Losing-Set Expansion}
A box $B$ may be certified losing under the continuation alphabet if:
\begin{equation}
    \forall a\in\mathcal  A_{\text{cont}},\;
    \forall B'\in\Succ(B,a):
    \quad B'\subseteq\underline{\mathcal L}.
\end{equation}

Then $B\subseteq\mathcal L^*$ and it may be added to
$\underline{\mathcal L}$. If we find a losing box, our implementation greedily attempts to certify the whole policy cell that contains the certified losing box, as losing.

\paragraph{Joint Realization.}
\textsc{Joint} searches for a controller $\pi'\in\Pi_R$ such that
\(
    \pi' \models \Gamma\cup\overline\Gamma .
\) $\Gamma$ denotes the \emph{committed} constraints from prior repair rounds,
while $\overline\Gamma$ represents the \emph{current} constraints that were
generated under the current $\pi'$. As all constraints are derived
independently they need not be jointly realizable, since each policy cell must
be assigned a single action. We encode their joint realization as a CP-SAT
problem using OR-Tools~\citep{cpsatlp}. Binary variables $w_{q,a}$ select one
action per cell, and $y_v$ indicates activation of a game node $v$:
\begin{equation}
\sum_{a\in\mathcal A(q)} w_{q,a}=1
\quad \forall q\in\mathcal Q,
\qquad\qquad
y_{\mathrm{root}(\gamma)}=1
\quad \forall \gamma\in\Gamma\cup\overline\Gamma,
\end{equation}
where $\mathrm{root}(\gamma)$ denotes the root node $v_0$ of the game DAG
$\mathcal G$ of $\gamma=(\mathcal G,i)$.

Every active node must select a stored safe action, and selecting an action
activates all corresponding child nodes: for every node $v$ of a game in
$\Gamma\cup\overline\Gamma$,
\begin{equation}
\begin{aligned}
y_v &\implies \bigvee_{a\in\mathcal A_v} w_{q(B_v),a},\\
y_v\land w_{q(B_v),a} &\implies y_{v'}
\quad\text{for every $a\in\mathcal A_v$ and every child $v'$ of $v$ induced by $a$}.
\end{aligned}
\end{equation}

\paragraph{Joint-Realization Heuristics.}
Optimal joint realization can be expensive, so \textsc{Joint} employs several
heuristics. For a set of repair constraints $\Sigma$, define its
\emph{support} as the set of policy cells occurring in any of its game DAGs,
\begin{equation}
    \operatorname{supp}(\Sigma)
    :=
    \left\{
        q(B_v)
        \;\middle|\;
        \gamma=(\mathcal G,i)\in\Sigma,\;
        v\text{ is a node of }\mathcal G
    \right\}.
\end{equation}
We call $\operatorname{supp}(\overline{\Gamma})$ the
\emph{current constraint cells}.

\textsc{Joint} first restricts the realization search to the current
constraint cells, such that policy cells outside
$\operatorname{supp}(\overline{\Gamma})$ are not modified. If no feasible
joint realization is found, the search is repeated over
\[
    \operatorname{supp}(\Gamma\cup\overline{\Gamma}),
\]
allowing cells occurring in both committed and current constraints to be
modified.

Among feasible assignments, \textsc{Joint} searches for one minimizing the
distance to the original controller. We limit this optimization to
$60\,\mathrm{s}$ and, if optimality has not been established by then, accept
the best feasible assignment found. %

\paragraph{Additional implementation details.} The original BBReach implementation uses Ariadne~\citep{benvenuti2008reachability} for post-image propagation. We instead implement propagation directly on the GPU using interval arithmetic for $\overlinePost$ and sound continuous-time bounds for $\overlineTube$. For efficiency, boxes in the same policy cell are merged by their bounding box. This is a more aggressive merging scheme than employed in~\citep{tian2023taming}. Benchmark-specific propagators are given in Appendix~\ref{app:environments}. We perform each tube check for $10$ segments, escalating to $100$ for any tubes where violations were found. Only after this refinement do we accept a tube as failing. Furthermore, after a certified repair is found, the repair is localized to only cells visited by the repaired controller. This information can be simply extracted from the stored forward propagation sets. Last, floating-point operations are soundly overapproximated or checked for exactness throughout the implementation.

\section{Appendix: Theorems}
\label{app:proofs}

\paragraph{Verifier assumptions.}

We build on the sound reachability procedure of
\citet{tian2023taming}.
Our proofs rely on the following assumptions:

\begin{enumerate}
\item \label{as:reach} \textsc{ReachStep} is covering: for every accepted step $k$ and every
$B \in \overline{\mathcal{X}}_k$,
$\Post(B, \pi'(q(B))) \subseteq \bigcup \overline{\mathcal{X}}_{k+1}$,
where $\overline{\Post}$ and $\overline{\Tube}$ are sound.
\item \label{as:discjoint} \textsc{ReachStep} reports every source box whose abstract tube intersects
$\mathcal U$ or whose abstract post-image intersects $\underline{\mathcal L}$. Hence,
\[
    \mathcal F=\emptyset
    \;\Longrightarrow\;
    \forall B\in \overline{\mathcal{X}}_k:\ \mathrm{SafeStep}(B,\pi'(q(B))).
\]

\item \label{as:earliest}\textsc{EarliestAffected} returns a rewind time step no later than any
previously verified step affected by a controller modification, an expansion
of $\underline{\mathcal L}$, or a refinement of the reachability computation with respect to merging;
the subsequent sets are recomputed.

\item \label{as:backtrack} Every obligation $(B,h)$ in the backward cone
(the initial obligations of Eq.~\ref{eq:frontier} and every obligation
produced from them by \textsc{BackTrack}) satisfies $\neg\mathrm{Safe}_h(B,\pi)$,
where $\pi$ is the controller that generated the failing frontier.

\item \label{as:cont} \textsc{FindCont} evaluates candidate actions with the same
$\mathrm{SafeStep}$ predicate and $\mathrm{Succ}$ relation as the forward
verifier: every action $a$ stored at a game node $v$ satisfies
$\mathrm{SafeStep}(B_v,a)$, and, if $h_v>1$, \emph{all} boxes in
$\mathrm{Succ}(B_v,a)$ appear as children of $v$ with horizon $h_v-1$.

\item \label{as:joint}  All calls to \textsc{ReachStep}, \textsc{Joint}, \textsc{FindCont}, \textsc{EarliestAffected}, \textsc{Backtrack} terminate. In particular \textsc{Joint} either returns \textsc{Unknown}, \textsc{SAT} with a valid repair, or \textsc{Backtrack}, with $\mathcal F_o \neq \emptyset$.
\item \label{as:finite} Assume further that
each abstract successor and predecessor computation returns finitely many
boxes and that the continuation backend $\mathcal W$ terminates. 
\item  \label{as:goal} For reach–avoid tasks we treat the goal set $\mathcal{X_G}$ as terminal: a trajectory that enters  $\mathcal{X_G}$ (satisfying the safety condition at entry) has completed the task and is not modeled further. Certified safety then means: every closed-loop trajectory from $\mathcal X_0$ avoids $\mathcal U$ until it enters $\mathcal{X_G}$. Non-entering trajectories avoid $\mathcal U$ till $T$.
\end{enumerate}

For simplicity, we state the guarantees with respect to the unmerged
reachability computation. Merging is used only as a scalability optimization
in the implementation. Whenever a merged box contributes to a failing or
unresolved frontier, \textsc{Backtrack} replays the corresponding reachability
step without merging and future forward steps retain this refinement. Thus all repair constraints
are derived from the unmerged predecessor relation.

\thmsoundness*
\begin{proof}
By assumption~\ref{as:reach}, every accepted reachability step
soundly covers the sampled nonterminal successors, and its inter-sample
tubes are disjoint from $\mathcal U$ by
assumption~\ref{as:discjoint}.
Whenever $\pi'$, $\underline{\mathcal L}$, or the reachability
computation changes, \textsc{EarliestAffected} causes every
potentially invalidated step to be recomputed
(assumption~\ref{as:earliest}).
Thus, at termination, all steps in the accepted prefix are
valid for the returned controller $\pi'$.

If the algorithm returns $(\pi',T)$, every step
$j=0,\ldots,T-1$ has been accepted. Hence,
\[
  \overlineReach_{\pi',T}(\mathcal X_0)
  \cap\mathcal U=\emptyset,
\]
so $\pi'$ is certified safe to horizon $T$ by
Definition~\ref{def:certified-safety}.

If the algorithm instead returns $(\pi',\infty)$ at step $k$,
every step $j=0,\ldots,k$ has been accepted, and the
fixed-point test establishes
\[
  \bigcup\overline{\mathcal X}_{k+1}
  \subseteq
  \bigcup\overline{\mathcal X}_k.
\]
Together with
\[
  \overlineReach_{\pi',k+1}(\mathcal X_0)
  \cap\mathcal U=\emptyset,
\]
this establishes certified safety to horizon $\infty$
by Definition~\ref{def:certified-safety}.

By assumption~\ref{as:reach}, the computed sets and tubes
cover the concrete trajectories throughout the accepted
prefix. In the fixed-point case, assumption~\ref{as:reach}  and
the inclusion above allow the accepted transition from
$\bigcup\overline{\mathcal X}_k$ to be applied inductively.
Its tube checks therefore establish safety for all
subsequent steps (assumption~\ref{as:discjoint}).

For reach-avoid tasks, this argument applies to trajectories
that have not terminated: every successor outside
$\mathcal X_G$ remains covered by the propagated sets.
Trajectories removed upon safe goal entry have completed
the task under assumption~\ref{as:goal}.
Thus concrete trajectories avoid $\mathcal U$ throughout
the certified horizon, subject to this terminal-goal
convention.
\end{proof}

\begin{lemma}[Failure constraints eliminate the current controller]
\label{lemma:constraint-removes-failure}
Let $\pi\in\Pi_R$ be the controller whose forward propagation generated a failure,
and let $\gamma=(\mathcal G,i)$ be a repair constraint generated
from a backward obligation $(B,h)$, where $\mathcal G$ has horizon $h$. Then
\[
    \pi \not\models \gamma .
\]
Consequently, for every $\Gamma$ such that $\pi\models\Gamma$,
\[
    \Pi(\Gamma\cup\{\gamma\})
    \subsetneq
    \Pi(\Gamma),
\]
where
\[
    \Pi(\Sigma)
    :=
    \{\pi'\in\Pi_R\mid \pi'\models\Sigma\}.
\]
\end{lemma}

\begin{proof}
Throughout this proof, $\mathrm{Safe}_h$ is evaluated with the
losing set used when the obligation was established.
Since the losing set only grows, every subsequent
continuation-game check also enforces this predicate. $(B,h)$ lies in the backward cone of a failure of $\pi$ (assumption \ref{as:backtrack}),
\[
    \neg\mathrm{Safe}_h(B,\pi).
\]
By soundness of the continuation-game constraint (assumption \ref{as:cont}),
\[
    \pi'\models\gamma
    \;\Longrightarrow\;
    \mathrm{Safe}_h(B,\pi').
\]
Hence
\[
    \pi\not\models\gamma.
\]
If additionally $\pi\models\Gamma$, then
\[
    \pi\in
    \Pi(\Gamma)\setminus\Pi(\Gamma\cup\{\gamma\}),
\]
and therefore
\[
    \Pi(\Gamma\cup\{\gamma\})
    \subsetneq
    \Pi(\Gamma).
\]
\end{proof}
\begin{lemma}[Termination of backward repair]
\label{lemma:brepair-termination}
Assume a finite verification horizon $T$, finitely branching predecessor
expansion, a terminating continuation backend $\mathcal W$, and finite
committed constraint sets. Then every call to \textsc{BRepair} terminates.
\end{lemma}
\begin{proof} Persistent merge refinement and its associated replay introduce
only finitely many additional steps; it therefore suffices to
consider a suffix without such steps.
Consider an arbitrary call to \textsc{BRepair} caused by a failure at
time step $ 0\le t< T$. Suppose, for contradiction, that the call has an infinite
execution trace.

Let $\mathcal T_t$ be the backward unfolding of the failing frontier.
Its nodes are occurrences of box-horizon pairs $(B,h)$ and
\[
    (B,h)\rightarrow(B',h+1),
    \qquad
    B'\in\textsc{Backtrack}(B).
\]
The initial frontier is finite, every node has finite out-degree (assumption~\ref{as:finite}), and
\[
    1\le h\le t + 1 \le T.
\]
Hence
\[
    |\mathcal T_t|<\infty.
\]

Consider an arbitrary iteration of the assumed infinite trace.
Since the trace does not terminate, neither \textsc{BRepair} nor
\textsc{Joint} returns \textsc{Repair} or \textsc{Unknown}~(assumption \ref{as:joint}).

Thus the iteration can continue only by
\begin{enumerate}
    \item returning one or more current constraints for further
          backtracking, or
    \item removing one or more previously committed non-anchored constraints
          in favor of an anchored current constraint.
\end{enumerate}

In Case~1, each returned pair $(B,h)$ is replaced by predecessors
$(B',h+1)$. Hence the corresponding occurrence advances strictly in the
finite tree $\mathcal T_t$. A current constraint that is returned for
backtracking is removed from $\overline\Gamma$ and therefore cannot be
backtracked again at the same occurrence. Thus Case~1 can occur only
finitely many times.

In Case~2, at least one constraint from the committed set present at entry
to the call is removed. Since this set is finite, Case~2 can also occur only finitely
many times.

Hence the assumed infinite trace cannot exist. Therefore every call to
\textsc{BRepair} terminates.
\end{proof}

\thmtermination*
\begin{proof}
By Lemma \ref{lemma:brepair-termination}, every call to \textsc{BRepair} terminates.
It remains to show that only finitely many such calls can occur.

Since $\mathcal Q$ and $\mathcal A_{\cont}$ are finite,
\[
    |\Pi_R|\le (|\mathcal A_{\cont}|+1)^{|\mathcal Q|}<\infty.
\]

We apply the following argument between losing-set updates,
merge refinements, and their associated replay steps, starting
after the first successful realization in such an interval.

Let $\Gamma_r$ be the committed constraint set immediately before the
$r$-th call to \textsc{BRepair}, let
\[
    A_r
    :=
    \{\gamma\in\Gamma_r\mid \gamma\text{ is anchored}\},
\]
and let $\pi_r$ be the current controller. By construction of
\textsc{Joint} (assumption \ref{as:joint}, trivial for $r=0$),
\begin{equation}
    \label{eq:modelsconst}
    \pi_r\models\Gamma_r.
\end{equation}

Consider an arbitrary successful call
$r$ in this sequence.

There are two cases.

\emph{Case 1: no committed constraint is removed.} Then
\[
    \Gamma_r\subseteq\Gamma_{r+1},
    \qquad\text{and hence}\qquad
    \Pi(A_{r+1})\subseteq\Pi(A_r).
\]
At least one constraint $\gamma$ generated from the current failure is newly
committed. By Lemma~\ref{lemma:constraint-removes-failure},
\[
    \pi_r\not\models\gamma.
\]
Together with~\ref{eq:modelsconst},
\[
    \pi_r\in
    \Pi(\Gamma_r)\setminus\Pi(\Gamma_{r+1}),
\]
and therefore
\begin{equation}
\label{eq:strictly-decreasing}
    \Pi(\Gamma_{r+1})
    \subsetneq
    \Pi(\Gamma_r).
\end{equation}

\emph{Case 2: a committed constraint is removed.}
By construction, this is permitted only in favor of a newly generated
anchored constraint $\gamma$; anchored constraints are never discarded. Hence
\[
    A_r\subsetneq A_{r+1}.
\]
Again, by Lemma~\ref{lemma:constraint-removes-failure},
\[
    \pi_r\not\models\gamma.
\]
Since $A_r\subseteq\Gamma_r$, \eqref{eq:modelsconst} gives $\pi_r\models A_r$, so
\[
    \pi_r\in
    \Pi(A_r)\setminus\Pi(A_{r+1}),
\]
and therefore
\begin{equation}
    \Pi(A_{r+1})
    \subsetneq
    \Pi(A_r).
    \label{eq:anchored-decrease}
\end{equation}

By \eqref{eq:anchored-decrease} in Case~2, and by
\eqref{eq:strictly-decreasing} together with $\Pi(A_{r+1})\subseteq\Pi(A_r)$
in Case~1, every successful call strictly decreases
\[
    \left(
        |\Pi(A_r)|,\,
        |\Pi(\Gamma_r)|
    \right)
\]
in lexicographic order. Since
\[
    \Pi(A_r),\Pi(\Gamma_r)\subseteq\Pi_R
\]
and $|\Pi_R|<\infty$, only finitely many calls to
\textsc{BRepair} can occur while $\underline{\mathcal L}$ is fixed.

Thus only finitely many successful repair calls can occur
in such an interval.

The merge mask has finitely many timestep--cell entries and
grows monotonically. Moreover, fixed, finitely branching
propagation from the finite initial representation, over the
finite action alphabet and finite horizon, generates only
finitely many candidate losing boxes.
Hence the losing set can also grow strictly only finitely
many times.

Each refinement entails only finitely many replay
steps. Consequently, there are only
finitely many such intervals and repair calls.
Between repairs and updates, forward verification advances
over the finite horizon. Hence \method terminates.

\end{proof}

\thmcompleteness*
\begin{proof}
Since $\mathcal A_{\cont}=\mathcal A$, we have
$\Pi_R=\mathcal A^{\mathcal Q}$.
By monotone losing-set expansion,
$\underline{\mathcal L}\subseteq L'$ throughout execution.
Therefore
\[
    \mathrm{Safe}^{L'}_h(B,\pi)
    \;\Longrightarrow\;
    \mathrm{Safe}^{\underline{\mathcal L}}_h(B,\pi).
\]
Thus the witness passes every safety test used during
construction of the exhaustive unmerged games.

Let $G(B,h)$ be the exhaustive continuation game. For $\pi\in\Pi_R$,
equation~\ref{eq:safe-h} and induction on $h$ give
\[
    \mathrm{Safe}_h^{L'}(B,\pi)\;\Longrightarrow\;\pi\models G(B,h):
\]
the action $\pi(q(B))$ passes $\mathrm{SafeStep}$, every successor admits
its continuation under $\pi$, and, since the exhaustive backend explores
every action of $\mathcal A$, the game retains this action and all its
successor branches at every node (assumption \ref{as:cont}).

Suppose, for contradiction, that the algorithm returns
\textsc{Unknown}.

Every anchored constraint $\gamma=(\mathcal G,0)$ has $B_{v_0}\in\mathcal X_0$ and
remaining horizon $h\le T$ (obligations never exceed horizon $t+1\le T$,
cf.\ line~15 of Algorithm~\ref{alg:backward-repair}). Due to the replay induced by merging the anchored constraint is entirely unmerged. Since $\mathrm{Safe}_T^{L'}$ implies
$\mathrm{Safe}_h^{L'}$ for $h\le T$, the witness satisfies
$\mathrm{Safe}_h^{L'}(B_{v_0},\overline\pi)$ and hence $\gamma$. This holds for both
committed and current constraints, so
\[
    \overline\pi\models
    \{\gamma\in\Gamma\cup\overline\Gamma:\gamma\text{ is anchored}\}.
\]
Thus no obligation at time step zero lacks a winning continuation, and no
anchored subset is jointly infeasible. 
Hence the algorithm cannot return \textsc{Unknown}.

By Theorem~\ref{thm:termination} it terminates and returns
some $(\pi',H)$ with $H\in\{T,\infty\}$; by Theorem~\ref{thm:soundness},
$\pi'$ is certified safe to horizon $H$, in particular safe for the first
$T$ control steps.

\end{proof}

\corbackendcompleteness*
\begin{proof}
The proof of Theorem \ref{thm:relative-completeness} applies unchanged, since the backend explores the witness
action $\overline\pi(q(B))$ at every relevant node.
\end{proof}
\section{Appendix: Environments}
\label{app:environments}

\subsection{Linear ACC}
\paragraph{System Dynamics.}
Our implementation follows \citep{teuber2024provably}.
Let $p_t$ denote the distance from the controlled vehicle to the lead
vehicle and $v_t$ their relative velocity, with $v_t<0$ indicating
that the vehicles are approaching each other. Under a relative
acceleration $a_t$ held constant over one sampling interval,
the dynamics are
\[
\begin{bmatrix}
p_{t+1} \\
v_{t+1}
\end{bmatrix}
=
\begin{bmatrix}
1 & \Delta t \\
0 & 1
\end{bmatrix}
\begin{bmatrix}
p_t \\
v_t
\end{bmatrix}
+
\begin{bmatrix}
\tfrac12(\Delta t)^2 \\
\Delta t
\end{bmatrix}
a_t.
\]
We use a sampling period of $\Delta t=0.1\,\mathrm{s}$ and
relative acceleration bounds
$a_t\in[-100,100]\,\mathrm{m/s^2}$.
The normalized controller action $u_t\in[-1,1]$ determines
$a_t=100u_t\,\mathrm{m/s^2}$; positive actions correspond
to ego-vehicle braking.

We consider collision avoidance, requiring $p(\tau)>0$
at all times, including between samples.
For this specification, the analytic braking controlled-invariant
region is
\[
\mathcal C_{\mathrm{brake}}
=
\left\{
(p,v):
p>\frac{\max(-v,0)^2}{2a_{\max}}
\right\},
\qquad a_{\max}=100\,\mathrm{m/s^2}.
\]
The condition ensures that the available distance exceeds
the stopping distance under maximal braking when the
vehicles are approaching each other.

\paragraph{Tube over-approximation.}
Under a constant relative acceleration $a$, the continuous-time
trajectory from an initial state $(p_0,v_0)$ is
\[
p(\tau)=p_0+v_0\tau+\tfrac12 a\tau^2,
\qquad
v(\tau)=v_0+a\tau,
\qquad \tau\in[0,\Delta t].
\]
For an initial box
$B=[\underline p,\overline p]\times[\underline v,\overline v]$,
position is monotone in both initial coordinates for $\tau\geq0$.
Consequently, its lower and upper bounds over a time segment
$[\tau_0,\tau_1]$ are
\[
\underline p_{\Tube}
=\min_{\tau\in[\tau_0,\tau_1]}
 \left(\underline p+\underline v\tau+\tfrac12 a\tau^2\right),
\qquad
\overline p_{\Tube}
=\max_{\tau\in[\tau_0,\tau_1]}
 \left(\overline p+\overline v\tau+\tfrac12 a\tau^2\right).
\]
Each extremum is obtained by evaluating the corresponding quadratic
at the segment endpoints and, when $a\neq0$, at its stationary
point $\tau_*=-v_0/a$ if that point lies within the segment.
Velocity bounds follow from the endpoint extrema of
$\underline v+a\tau$ and $\overline v+a\tau$.
Together, these bounds form an axis-aligned enclosure of every
trajectory starting in $B$ throughout the segment.
\paragraph{Reinforcement Learning.} Again following \citep{teuber2024provably}, all controllers are trained with PPO~\citep{schulman2017proximal} for a total of 200,000 time steps. In contrast to~\citet{teuber2024provably}, we train with cleanrl~\citep{huang2022cleanrl} instead of stable-baselines-3~\citep{stable-baselines3}. Our five seeds results are reported in Table \ref{tab:acc-evaluation}. We utilize the hyperparameters given in Table \ref{tab:training}.

\begin{table}[h]
\centering
\small
\caption{ACC: training hyperparameters.}
\label{tab:training}
\begin{tabular}{llr}
\toprule
& Hyperparameter & Value \\
\midrule
\multirow{12}{*}{\rotatebox[origin=c]{90}{PPO}}
& Total time steps                     & 200000 \\
& Learning rate                       & $3\cdot10^{-4}$ (annealed) \\
& Parallel environments               & 1 \\
& Rollout steps                       & 2048 \\
& Batch      & 2048 \\
& Update epochs                       & 10 \\
& Discount $\gamma$                   & 0.99 \\
& GAE $\lambda$                       & 0.95 \\
& Clipping coefficient                & 0.2 \\
& Clipped value loss                  & true \\
& Advantage normalization             & true \\
& Entropy / value-loss coefficient    & 0.0 / 0.5 \\
& Max gradient norm, target KL        & 0.5, none \\
\midrule
\multirow{8}{*}{\rotatebox[origin=c]{90}{DWC}}
& Thermometer bits per dimension      & {81}  \\
& Thermometer type                    & {uniform} \\
& Grid domain                         & $[-2.5, 202.5]$\,m, $\pm205$\,m/s \\
& LUT layers $\times$ width           & {$2\times128$} \\
& LUT fan-in                          & {6} \\
& Initial $\log\alpha$                & $-0.6931$ \\
& Final \texttt{tanh} squashing       & off \\
& Critic layers $\times$ width, act.  & {$4\times64$}, {ReLU} \\
\bottomrule \\
\end{tabular}
\end{table}

For the ReLU baseline, we use the 4-layer MLP from \citep{teuber2024provably} with 64 units per layer, and for the DWC baseline we use a 2-layer DWC with 128 LUT-6 per layer. The per-step reward function is defined as $r_t = 10 (1 - |u|)$, encouraging small action changes. In addition, a large penalty of $-2000$ is applied for collisions. When the distance between the cars grows larger than $ p_{rel} > 200$, a penalty of $-800$ is given ending the episode.

\subsubsection{Baselines}

In Table~\ref{tab:acc-runtime}, Change [\%] denotes the fraction of states for which the repaired controller output differs from the original controller output by more than \(10^{-6}\). For ReLU networks, this quantity is estimated using \(100{,}000\) Monte Carlo samples. Mean \(\Delta\) denotes the average magnitude of the controller-output change over all sampled states. 

\paragraph{BBReach + Supervised} \begin{wrapfigure}{r}{0.5\textwidth}
\centering
\pgfplotscreateplotcyclelist{myBoxplotColors}{
  {fill=blue!70!black},
  {fill=orange!85!black},
  {fill=red!75!black},
  {fill=violet!80!black},
  {fill=teal!70!black},
  {fill=magenta!70!black},
  {fill=gray!70!black},
  {fill=brown!70!black}
}
\begin{tikzpicture}
\begin{semilogyaxis}[
  width=\linewidth, height=4.1cm,
  font=\footnotesize,
  ybar, bar width=4.5pt,
  cycle list name=myBoxplotColors,
  ymin=0.1, log origin y=infty,
  enlarge x limits=0.22,
  symbolic x coords={g0,g1,g4},
  xtick={g0,g1,g4},
  xticklabels={\shortstack{No losing\\set},\shortstack{Online\\losing set},Invariant},
  xticklabel style={align=center},
  ylabel={Games played},
  ytick={1,10,100,1000},
  legend style={at={(0.5,1.03)},anchor=south,legend columns=4,
                font=\scriptsize,/tikz/every even column/.append style={column sep=3pt}},
  error bars/y dir=both, error bars/y explicit,
  error bars/error bar style={black, line width=0.5pt},
  error bars/error mark options={rotate=90,mark size=1.6pt},
]
  \addplot coordinates {
    (g0,575.6) += (0,297.264697) -= (0,297.264697)
    (g1,256.4) += (0,101.623324) -= (0,101.623324)
    (g4,83.4) += (0,44.3147831) -= (0,44.3147831)
  };
  \addlegendentry{Depth 1}
  \addplot coordinates {
    (g0,572.8) += (0,320.907619) -= (0,320.907619)
    (g1,198.8) += (0,98.4616677) -= (0,98.4616677)
  };
  \addlegendentry{Depth 2}
  \addplot coordinates {
    (g0,376.4) += (0,231.1705) -= (0,231.1705)
    (g1,103.4) += (0,40.960957) -= (0,40.960957)
  };
  \addlegendentry{Depth 3}
  \addplot coordinates {
    (g0,187.4) += (0,113.111892) -= (0,113.111892)
    (g1,54) += (0,22.5610283) -= (0,22.5610283)
  };
  \addlegendentry{Depth 4}
  \addplot coordinates {
    (g0,77.4) += (0,44.8586669) -= (0,44.8586669)
    (g1,19.8) += (0,13.0843418) -= (0,13.0843418)
  };
  \addlegendentry{Depth 5}
  \addplot coordinates {
    (g0,26) += (0,14.6116392) -= (0,14.6116392)
    (g1,8.2) += (0,7.49666593) -= (0,7.49666593)
  };
  \addlegendentry{Depth 6}
  \addplot coordinates {
    (g0,5.4) += (0,3.57770876) -= (0,3.57770876)
    (g1,0.4) += (0,0.547722558) -= (0,0.3)
  };
  \addlegendentry{Depth 7}
\end{semilogyaxis}
\end{tikzpicture}
\caption{ACC game depths for different losing set configurations.}
\label{fig:game_depths}
\end{wrapfigure}
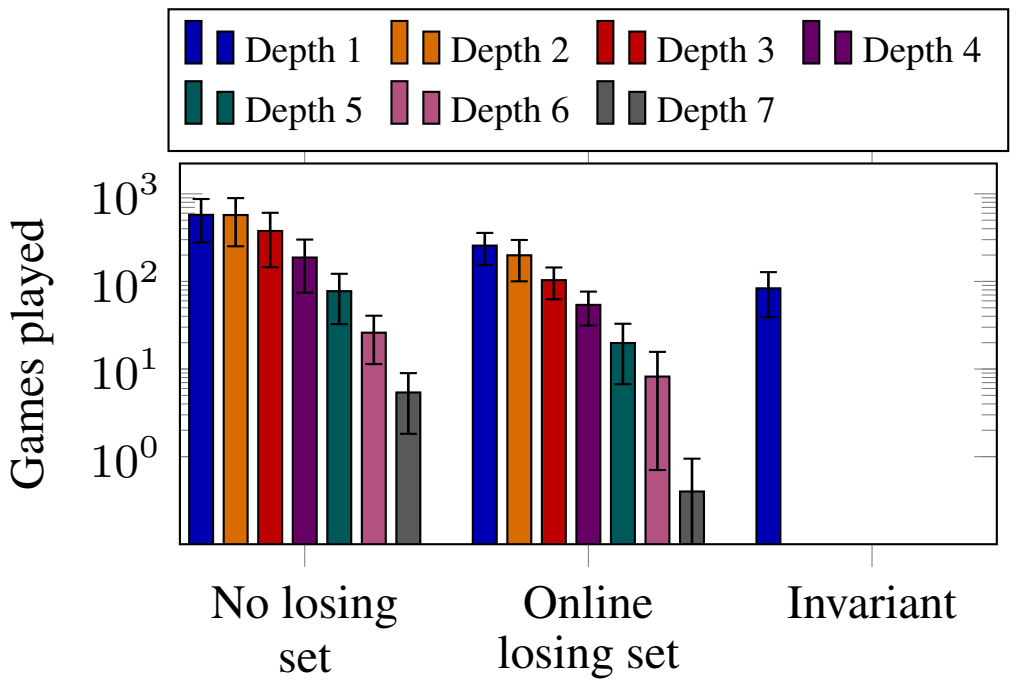

This baseline adapts the repair procedure of~\citet{yang2022neural}. We first perform multi-step closed-loop reachability analysis (\citet{yang2022neural} perform only one-step analysis) using our BBReach implementation. Whenever verification reaches the unsafe set, we collect the failing one-step transition and add the predecessor states to a supervised repair dataset, assigning a target action of \(1.05\), corresponding to full braking. To mitigate degradation of the original policy, we additionally collect \(100{,}000\) safe transitions generated by the unrepaired DWC and retain their original actions as supervised targets.

We then retrain the controller for \(1000\) gradient steps with batch size \(256\), sampling each minibatch equally from the repair and preservation datasets. After retraining, we rerun closed-loop reachability analysis and repeat this procedure until the controller verifies successfully for a horizon of \(50\) time steps. Finally, we run the ReachECO verifier requiring that no further repair constraints are generated. We report the total runtime of this complete repair-and-reverification loop.

The invariant is required as only one-step repair can be performed with this technique.
\paragraph{STLGym fine-tuning.}
\definecolor{seedzero}{RGB}{0,114,178}
\definecolor{seedone}{RGB}{213,94,0}
\definecolor{seedtwo}{RGB}{0,158,115}
\definecolor{seedthree}{RGB}{204,121,167}
\definecolor{seedfour}{RGB}{230,159,0}
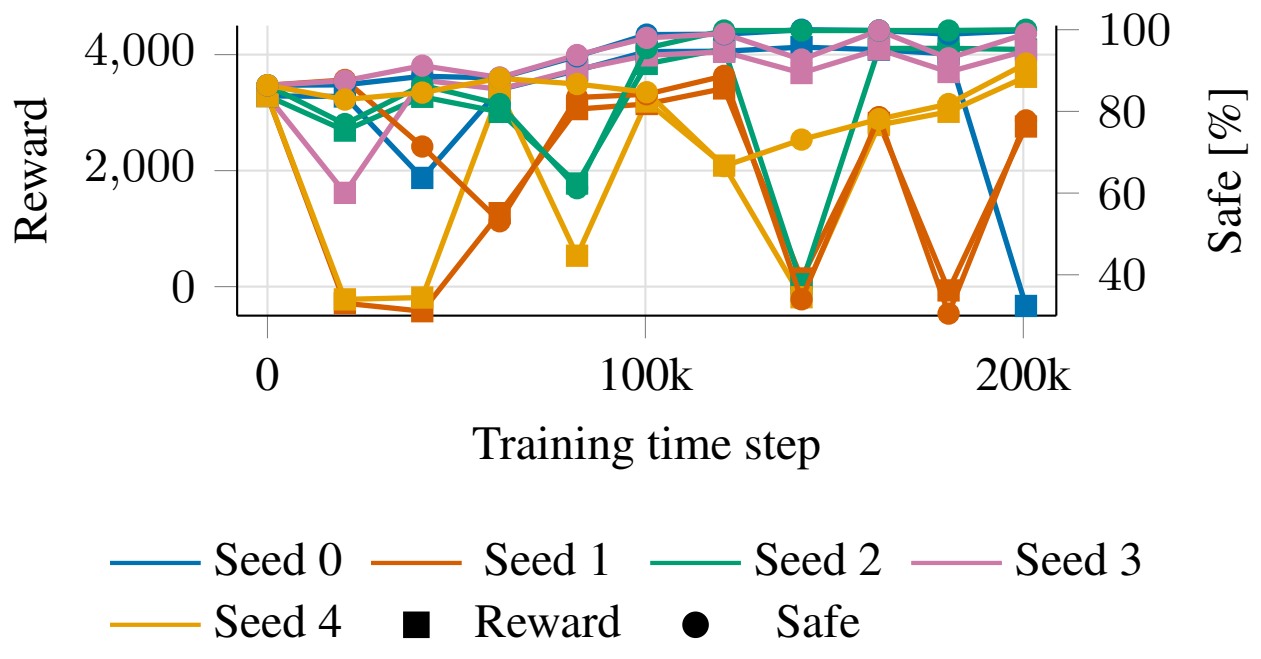
\begin{figure*}[t!]
\centering
\definecolor{seedzero}{RGB}{0,114,178}
\definecolor{seedone}{RGB}{213,94,0}
\definecolor{seedtwo}{RGB}{0,158,115}
\definecolor{seedthree}{RGB}{204,121,167}
\definecolor{seedfour}{RGB}{230,159,0}
\begin{tikzpicture}
\begin{axis}[
  width=7cm,
  height=3.5cm,
  xmin=-8028.16, xmax=208732.16,
  ymin=-500, ymax=4500,
  xlabel={Training time step},
  ylabel={Reward},
  xtick={0,100000,200000},
  xticklabels={0,100k,200k},
  minor tick num=0,
  axis y line*=left,
  axis x line*=bottom,
  tick align=outside,
  scaled x ticks=false,
  grid=both,
  grid style={black!12},
  font=\footnotesize,
  every axis plot/.append style={line width=1pt},
  mark size=1.6pt,
  legend to name=acctrainlegend,
  legend style={legend columns=4, font=\footnotesize, draw=none,
                /tikz/every even column/.append style={column sep=3pt}},
]
\addplot[color=seedzero, mark=square*, forget plot]
  coordinates {(0,3287.66501405) (20480,3278.74388842) (40960,1874.12316883) (61440,3398.00913007) (81920,3726.99766745) (100352,4052.58258296) (120832,4058.36463351) (141312,4128.96379244) (161792,4084.71478911) (180224,4011.98877715) (200704,-333.227874882)};
\addplot[color=seedone, mark=square*, forget plot]
  coordinates {(0,3287.66501405) (20480,-285.765786273) (40960,-423.773341479) (61440,1266.71214931) (81920,3060.74832303) (100352,3148.95276937) (120832,3417.06533813) (141312,141.153519891) (161792,2823.13569717) (180224,-65.0378362942) (200704,2761.07052664)};
\addplot[color=seedtwo, mark=square*, forget plot]
  coordinates {(0,3287.66501405) (20480,2690.20265623) (40960,3273.01908542) (61440,3012.87824416) (81920,1778.21965761) (100352,3839.94883756) (120832,4101.22197027) (141312,39.5964653741) (161792,4105.45106869) (180224,4108.76226612) (200704,4088.78733693)};
\addplot[color=seedthree, mark=square*, forget plot]
  coordinates {(0,3287.66501405) (20480,1616.90260074) (40960,3560.1906389) (61440,3402.54149194) (81920,3744.16377196) (100352,3986.99920539) (120832,4047.70156633) (141312,3682.62169227) (161792,4091.14764089) (180224,3700.67026275) (200704,4067.10736886)};
\addplot[color=seedfour, mark=square*, forget plot]
  coordinates {(0,3287.66501405) (20480,-219.495745902) (40960,-187.75175595) (61440,3369.72646507) (81920,533.065280016) (100352,3191.5157093) (120832,2086.98099482) (141312,-183.067945446) (161792,2786.78766181) (180224,3014.34117062) (200704,3617.88381961)};
\addlegendimage{color=seedzero, thick}
\addlegendentry{Seed 0}
\addlegendimage{color=seedone, thick}
\addlegendentry{Seed 1}
\addlegendimage{color=seedtwo, thick}
\addlegendentry{Seed 2}
\addlegendimage{color=seedthree, thick}
\addlegendentry{Seed 3}
\addlegendimage{color=seedfour, thick}
\addlegendentry{Seed 4}
\addlegendimage{color=black, only marks, mark=square*}
\addlegendentry{Reward}
\addlegendimage{color=black, only marks, mark=*}
\addlegendentry{Safe}
\end{axis}
\begin{axis}[
  width=7cm,
  height=3.5cm,
  xmin=-8028.16, xmax=208732.16,
  ymin=30, ymax=101,
  ylabel={Safe [\%]},
  minor tick num=0,
  axis y line*=right,
  axis x line=none,
  xtick=\empty,
  tick align=outside,
  font=\footnotesize,
  every axis plot/.append style={line width=1pt},
  mark size=1.6pt,
]
\addplot[color=seedzero, mark=*, forget plot]
  coordinates {(0,86.4) (20480,86.5) (40960,88.6) (61440,88.2) (81920,93.5) (100352,98.8) (120832,98.9) (141312,100) (161792,99.8) (180224,98.9) (200704,99.8)};
\addplot[color=seedone, mark=*, forget plot]
  coordinates {(0,86.4) (20480,87.7) (40960,71.4) (61440,53.1) (81920,83.4) (100352,84.2) (120832,88.7) (141312,34) (161792,78.6) (180224,30.5) (200704,77.8)};
\addplot[color=seedtwo, mark=*, forget plot]
  coordinates {(0,86.4) (20480,76.9) (40960,86.4) (61440,81.8) (81920,61.2) (100352,95.5) (120832,99.8) (141312,99.8) (161792,99.8) (180224,99.8) (200704,100)};
\addplot[color=seedthree, mark=*, forget plot]
  coordinates {(0,86.4) (20480,87.5) (40960,91.2) (61440,88.3) (81920,93.8) (100352,98) (120832,99) (141312,92.7) (161792,99.8) (180224,93) (200704,99)};
\addplot[color=seedfour, mark=*, forget plot]
  coordinates {(0,86.4) (20480,82.9) (40960,84.6) (61440,88.1) (81920,86.7) (100352,84.7) (120832,66.6) (141312,73.1) (161792,78.1) (180224,81.8) (200704,91.7)};
\end{axis}
\node at (current bounding box.south) [anchor=north, yshift=-1mm]
  {\pgfplotslegendfromname{acctrainlegend}};
\end{tikzpicture}

\caption{ACC reward and empirical safety during STL-Gym fine-tuning across five seeds~\citep{hamilton2022training}. Both reward and empirical safety vary across seeds and training checkpoints. None of the checkpoints achieves
$100\%$ empirical safety when additionally evaluated
on near-boundary initial states.}
\label{fig:acc-stlgym-training}
\end{figure*}

We adapt the dense robustness reward of \citet{hamilton2022training}
to the ACC specification
\[
\phi=\mathbf{G}_{[0,420]}(p_{\mathrm{rel}}>0),
\]
evaluated at the sampled states of each episode. Starting from the same
pretrained NCubeV ReLU PPO controller, we perform five independent
fine-tuning runs with a budget of \(200{,}000\) environment steps each.
At step \(t\), PPO receives
\[
r_t=\frac{1}{P}\min_{0\leq k\leq t}p_{\mathrm{rel},k},
\]
where \(P=100\).
To expose the controller to difficult initial states, we draw 10\% of
training resets slightly inside the analytic braking boundary. 
We evaluate each run at ten checkpoints during training. Every checkpoint has
sampled trajectories from near-boundary starts that reach
\(p_{\mathrm{rel}}\leq0\). We therefore do not attempt a formal safety certificate.

The 5 seed runs are shown in Figure~\ref{fig:acc-stlgym-training}. 

\paragraph{NCubeV + REASSURE}
This baseline uses the NCubeV framework introduced by
\citet{teuber2024provably} together with their four-layer ReLU MLP.
NCubeV utilizes a provably safe control envelope derived using differential
dynamic logic~\citep{DDL} to reduce infinite-horizon closed-loop safety
verification to neural-network verification against the envelope.
REASSURE~\citep{fu2022reassure} repairs ReLU networks by constructing localized
patch networks around identified buggy regions.

We combine the two approaches as follows. NCubeV identifies regions in which
the controller violates the safe control envelope, and we provide these regions
to REASSURE together with the desired controller outputs. REASSURE then
constructs local patches enforcing the desired behavior. For ACC, full braking
provides such a safe target action for states inside the controllable region.

\paragraph{NCubeV + Supervised} This baseline is similar to BBReach + Supervised; NCubeV identifies unsafe regions from which the supervised repair dataset is generated (each sample having its correct output set to 1.05). Additionally, an equal number of safe transitions (100,000) is produced by the unrepaired neural network. With this data, the neural network is retrained until a loss threshold is reached. This procedure is repeated until NCubeV can not provide any unsafe regions and verifies the safety of the controller.

\subsubsection{Extended Experimental 
Evaluation}

Figure~\ref{fig:game_depths} shows the depth of the games played with $\mathcal P_{max} = 30$. Learning the losing set decreases the number of deeper games. Table~\ref{tab:acc-evaluation} shows performance of the 5 DWC seeds before and after invariant repair, as well as the ReLU baseline from \citet{teuber2024provably}. 

After repair, all seeds are certified and hence, there are no remaining unsafe empirical trajectories.

\begin{table}[t]
\centering
\caption{ACC evaluation over 1,000 rollouts per controller. Reward is mean $\pm$ standard deviation. Unsafe denotes sampled collisions. All controllers, including the reference and invariant-repaired DWCs, use the same initial-state sampling protocol.}
\label{tab:acc-evaluation}
\begin{tabular}{lrrl}
\toprule
Seed & Reward & Unsafe (\%) & Architecture \\
\midrule
\citet{teuber2024provably} & $3254.16 \pm 2001.49$ & 12.70 & ReLU \\
\midrule
\multicolumn{4}{l}{\textit{Before repair}} \\
1 & $2439.83 \pm 2589.29$ & 25.60 & DWC \\
2 & $1957.28 \pm 2778.07$ & 33.40 & DWC \\
3 & $3183.04 \pm 1929.58$ & 12.20 & DWC \\
4 & $2090.53 \pm 2646.71$ & 29.80 & DWC \\
5 & $2499.29 \pm 2387.39$ & 22.10 & DWC \\
\midrule
\multicolumn{4}{l}{\textit{After invariant repair}} \\
1 & $3943.31 \pm 40.14$ & 0.00 & DWC \\
2 & $3916.49 \pm 38.87$ & 0.00 & DWC \\
3 & $3895.33 \pm 39.61$ & 0.00 & DWC \\
4 & $3797.75 \pm 37.71$ & 0.00 & DWC \\
5 & $3754.34 \pm 48.64$ & 0.00 & DWC \\
\bottomrule
\end{tabular}
\end{table}

\subsection{Docking2D}

\paragraph{System Dynamics.}
Planar spacecraft docking is modeled using the linear
Clohessy-Wiltshire-Hill equations~\citep{clohessy1960terminal}.
Let $p_t=(x_t,y_t)$ denote the relative position and
$v_t=(v_{x,t},v_{y,t})$ the relative velocity of the spacecraft.
Under a thrust $u_t=(u_{x,t},u_{y,t})$ held constant over one sampling
interval, the dynamics can be written as
\[
\dot{x}_t = A x_t + B u_t,
\]
with state $x_t=(x_t,y_t,v_{x,t},v_{y,t})^\top$ and
\[
A =
\begin{bmatrix}
0 & 0 & 1 & 0 \\
0 & 0 & 0 & 1 \\
3n^2 & 0 & 0 & 2n \\
0 & 0 & -2n & 0
\end{bmatrix},
\qquad
B =
\begin{bmatrix}
0 & 0 \\
0 & 0 \\
1/m & 0 \\
0 & 1/m
\end{bmatrix}.
\]
As in prior work, we use a sampling period of $\Delta t=1\,\mathrm{s}$,
spacecraft mass $m=12\,\mathrm{kg}$, and orbital mean motion
$n=0.001027\,\mathrm{s}^{-1}$. The normalized controller action
$u_t\in[-1,1]^2$ determines the thrust components in newtons.
These dynamics and parameters follow
\citep{lopez2025archcomp25ainncs,ravaioli2022safe,mandal2024formally}.
\paragraph{Specification} We follow the AINNCS competition benchmark
\citep{lopez2025archcomp25ainncs}, which requires
\[
\|v(\tau)\|_2
\leq
v_{\mathrm{offset}}+2n\|p(\tau)\|_2,
\qquad
v_{\mathrm{offset}}=0.2\,\mathrm{m/s},
\]
at all times, including between controller samples.
We take as initial set
\[
\mathcal X_0
=
[70,120]^2\times[-0.28,0.28]^2
\]
which is larger than in \citep{lopez2025archcomp25ainncs}.

The terminal goal is specified following \citep{mandal2024formally}:
\[
\mathcal {X_G}=
\left\{
(x,y,v_x,v_y):
|x|\leq0.35\,\mathrm{m},\;
|y|\leq0.35\,\mathrm{m}
\right\}.
\]

We remove any partial-box entering the goal set that satisfies the velocity bound from further reachability propagation, considering it successfully docked. 

\paragraph{Tube over-approximation.}
To account for the continuous-time evolution between controller samples,
we augment the state with the constant control input,
\[
z=(x,y,v_x,v_y,u_x,u_y)^\top.
\]
Since $\dot u_x=\dot u_y=0$, the augmented dynamics are linear,
\[
\dot z = Kz,
\qquad
K =
\begin{bmatrix}
A & B \\
0 & 0
\end{bmatrix}.
\]
Hence, for a constant action over one sampling interval, the trajectory is
given exactly by
\[
z(\tau)=e^{K\tau}z_0,
\qquad
\tau\in[0,\Delta t].
\]
Thus, continuous-time propagation only requires evaluating the matrix
exponential of the constant $6\times6$ matrix $K$.

For an initial state box $B$, the augmented initial set is
\[
Z=B\times\{u\}
=
[\underline z,\overline z].
\]
At any fixed time $\tau$, let
\[
M(\tau)=e^{K\tau}.
\]
The exact axis-aligned enclosure of the propagated box is
\[
\underline z(\tau)
=
M^+(\tau)\underline z
+
M^-(\tau)\overline z,
\qquad
\overline z(\tau)
=
M^+(\tau)\overline z
+
M^-(\tau)\underline z,
\]
where $M^+=\max(M,0)$ and $M^-=\min(M,0)$ are taken
elementwise. Evaluating these bounds at $\tau=\Delta t$ and retaining
the first four coordinates gives the next-state box.
To conservatively enclose the continuous-time reachable tube between
controller samples, we partition $[0,\Delta t]$ into time segments,
following standard continuous-time over-approximation constructions
systems~\citep{althoff2007reachability}. For a segment
$[\tau_0,\tau_1]$, define

$$
c=\frac{\tau_0+\tau_1}{2},
\qquad
\delta=\frac{\tau_1-\tau_0}{2}.
$$

Let $[\underline z(c),\overline z(c)]$ denote the reachable box at the
segment midpoint. For any $\tau\in[\tau_0,\tau_1]$, the deviation from the
midpoint state can be conservatively bounded componentwise by

$$
r=
\left(e^{|K|\delta}-I\right)
\max\left(
|\underline z(c)|,
|\overline z(c)|
\right).
$$

Hence, the reachable tube over the entire segment is contained in

$$
[\underline z(c)-r,\;
 \overline z(c)+r].
$$

The first four coordinates therefore give an axis-aligned
over-approximation of all states reached from $B$ during the segment.

\paragraph{Reinforcement Learning.}
We utilize the environment and reward functions of
\citet{ravaioli2022safe}, and train three seeds for two million environment
steps each. Our initial state lies on the circle of radius $130\,\mathrm{m}$
around the dock, and evaluation is run from this same set.
Table~\ref{tab:training-docking2d} summarizes the hyperparameters used for training the DWC controller with PPO.

\begin{table}
\centering
\caption{Docking2D: training hyperparameters.}
\label{tab:training-docking2d}
\begin{tabular}{llr}
\toprule
& Hyperparameter & Value \\
\midrule
\multirow{13}{*}{\rotatebox[origin=c]{90}{PPO}}
& Total time steps                     & 2000000 \\
& Learning rate                       & $5\cdot10^{-5}$ (constant) \\
& Parallel environments               & 4 \\
& Rollout steps per environment       & 1000 \\
& Batch / minibatch                   & 4000 \\
& Update epochs                       & 30 \\
& Discount $\gamma$                   & 0.968559 \\
& GAE $\lambda$                       & 0.928544 \\
& Clipping coefficient                & 0.3 \\
& Clipped value loss                  & true \\
& Advantage normalization             & true \\
& Entropy / value-loss coefficient    & 0.0 / 1.0 \\
& Max gradient norm, target KL        & 0.5, 0.01 \\
\midrule
\multirow{10}{*}{\rotatebox[origin=c]{90}{DWC}}
& Thermometer bits per dimension      & {161} \\
& Thermometer type                    & {Gaussian} ($x,y$), {uniform} ($v_x,v_y$) \\
& Grid domain                         & $\pm125$\,m, $\pm0.5$\,m/s \\
& LUT layers $\times$ width           & {$2\times1024$} \\
& LUT fan-in                          & {6} \\
& Initial $\log\alpha$                & $-0.6931$ \\
& Final \texttt{tanh} squashing       & on \\
& Critic layers $\times$ width, act.  & {$2\times64$}, {tanh} \\
\bottomrule \\
\end{tabular}
\end{table}

\subsection{Car2D}
\label{sec:car2d}

\paragraph{System dynamics.}
The cars state is comprised of $s=(x,y,\theta,v)$, planar position,
heading, and speed. The controller outputs an acceleration and
turn command $u=(u_a,u_\omega)\in[-1,1]^2$. Each command is held constant
for $\Delta t=1\,\mathrm{s}$ and scaled to physical acceleration
$a=0.25u_a$ and turn rate $\omega=0.5u_\omega$. The continuous dynamics are the same as the NAV benchmark in AINNCS~\citep{lopez2025archcomp25ainncs}.
\[
\dot x=v\cos\theta,\qquad
\dot y=v\sin\theta,\qquad
\dot\theta=\omega,\qquad
\dot v=a.
\]

Hence, $\theta(\tau)=\theta_0+\omega\tau$ and
$v(\tau)=v_0+a\tau$, while
\[
\begin{aligned}
x(\tau)&=x_0+\int_0^\tau (v_0+at)
                  \cos(\theta_0+\omega t)\,dt,\\
y(\tau)&=y_0+\int_0^\tau (v_0+at)
                  \sin(\theta_0+\omega t)\,dt.
\end{aligned}
\]

\paragraph{Specification.}
The unsafe set consists
of contact with any obstacle or violation of $|\theta|\leq\pi$.
The two rectangular obstacles, expressed as Cartesian products of
$x$ and $y$ intervals, are
\[
\begin{split}
O_1&=[2.4,3.6]\times[1.9,3.1],\\
O_2&=[4.9,6.1]\times[-1,5.9].\\
\end{split}
\]
For the $1\,\mathrm{m}\times1\,\mathrm{m}$ initial-box experiment, we
verify
\[
\mathcal X_0=
[-0.53,0.47]\times[2.05,3.05]\times
[-0.35,-0.25]\times[0.94,0.98].
\]
The goal set is $x \geq 6$. The verification budget was set to $T=100$ controller steps. After a safe
transition, any portion of a post-image box contained in $ \mathcal {X_G}$
is removed from further propagation.
\paragraph{Tube over-approximation.}
For a fixed controller action, the closed-form flow described above is used
to compute an interval enclosure of the state at any fixed time
$\tau\in[0,\Delta t]$. Evaluating this enclosure at $\tau=\Delta t$ gives
the post-image box. To conservatively enclose the continuous-time trajectory between controller
samples, we partition $[0,\Delta t]$ into time segments
$[\tau_0,\tau_1]$. For each segment, we first evaluate the closed-form flow
at $\tau_0$, obtaining a box
\[
B(\tau_0)
=
X(\tau_0)\times Y(\tau_0)\times
\Theta(\tau_0)\times V(\tau_0).
\]
Since
\[
\dot x=v\cos\theta,
\qquad
\dot y=v\sin\theta,
\]
we have
\[
|\dot x|\leq |v|,
\qquad
|\dot y|\leq |v|.
\]
Let
\[
V_{\max}
=
\max_{\substack{
v_0\in[\underline v,\overline v]\\
\tau\in[\tau_0,\tau_1]
}}
|v_0+a\tau|.
\]
Because $v_0+a\tau$ is affine in both $v_0$ and $\tau$, this maximum is
attained at one of the four endpoint combinations. Writing
$h=\tau_1-\tau_0$, every trajectory over the segment therefore satisfies
\[
x(\tau)\in X(\tau_0)+[-hV_{\max},hV_{\max}],
\qquad
y(\tau)\in Y(\tau_0)+[-hV_{\max},hV_{\max}].
\]
Heading and speed are affine in time,
\[
\theta(\tau)=\theta_0+\omega\tau,
\qquad
v(\tau)=v_0+a\tau,
\]
so their bounds over $[\tau_0,\tau_1]$ are obtained directly from their
endpoint extrema. The Cartesian product of these four intervals forms a
sound axis-aligned over-approximation of the reachable tube over the
segment.

\paragraph{Controller training.}
The DWC controller was trained with TD3~\citep{fujimoto2018addressing} for $120{,}000$ environment steps. The training reward is progress
towards the environment's goal box, minus a per-step and action penalty,
with an additional reward for reaching the goal and a penalty for
unsafe termination. The trained policy uses a
$20\times20\times10\times10$ policy-grid.

\section{Circuit Size}
\label{app:dwc-size}

We use the DWC hardware synthesis pipeline of~\citet{kresse2026differentiable} and extend it with our ECO patch. We first briefly review the circuit of a DWC, then describe the patch circuit and report hardware numbers.

\subsection{Differentiable Weightless Controller circuit}

A DWC consists of four stages~\citep{kresse2026differentiable}, with the first three mirroring Differentiable Weightless Networks~\citep{bacellar2024dwn} and Logic Gate Neural Networks~\citep{petersen2022logic}:
\begin{enumerate}
  \item[(1)] \textbf{Thresholding.} Each observation, given as a 12-bit fixed-point word, is compared against the learned thresholds of its dimension. The comparator outputs form the thermometer encoding: bit~$j$ is set if and only if the observation exceeds threshold~$j$.
  \item[(2)] \textbf{LUT layers.} A stack of layers (two throughout this work), each a set of 6-input lookup tables. The inputs of every LUT are wired to fixed outputs of the previous layer, so each LUT maps directly onto one LUT6 primitive of the FPGA.
  \item[(3)] \textbf{Popcount.} Per action dimension, the outputs of the final layer assigned to it are summed by an adder tree.
  \item[(4)] \textbf{Output lookup.} A final table maps each popcount to an action value.
\end{enumerate}

We omit stage~(4): for typical DWCs it requires one BRAM per action dimension, and the patch affects it only marginally (it may require a larger memory, which is very unlikely to exceed the capacity of a single BRAM).

Throughout, we set the number of pipeline stages of the DWC controller to eight cycles (nine including the final BRAM). The patch contains a matching number of pipeline stages, so it adds no latency. Fewer pipeline stages are possible; we did not sweep this parameter. All synthesis results were obtained with Vivado~2022.2, with implementation on a Xilinx Artix-7 xc7a100t (speed grade~$-1$) targeting a clock frequency of 100\,MHz. All reported designs meet this target.

Table~\ref{tab:hardware} reports the synthesized sizes of the investigated DWCs.

\begin{table}[t]
\centering
\caption{Synthesized hardware at $\mathcal P_{max}=100$ for ACC (mean and standard deviation over five seeds) and for the Car2D and Docking2D controllers. \emph{Cells} is the number of repaired policy cells. The controller is the DWC without its patch (thresholding, LUT layers and popcounts with their pipeline registers). Vivado~2022.2, Artix-7 xc7a100t ($-1$); all designs meet 100\,MHz with and without their patch.}
\label{tab:hardware}
\small
\setlength{\tabcolsep}{4pt}
\begin{tabular}{llcccccc}
\toprule
& & & \multicolumn{2}{c}{Controller} & \multicolumn{3}{c}{Patch} \\
\cmidrule(lr){4-5} \cmidrule(lr){6-8}
& & Cells & LUT & FF & Logic LUT & Logic FF & Memory LUT \\
\midrule
ACC & $\mathcal L^*$ & $91_{\pm 22}$ & $1{,}273_{\pm 24}$ & $584_{\pm 4}$ & $114_{\pm 21}$ & $114_{\pm 23}$ & $15_{\pm 3}$ \\
ACC & $\mathcal W_{\mathrm{brake}}$ & $153_{\pm 14}$ & $1{,}273_{\pm 24}$ & $584_{\pm 4}$ & $182_{\pm 17}$ & $176_{\pm 15}$ & $28_{\pm 2}$ \\
\midrule
Car2D &  & 39 & 1{,}969 & 1{,}708 & 64 & 62 & 12 \\
\midrule
Docking2D & seed 1 & 5 & 6{,}860 & 3{,}873 & 36 & 43 & 14 \\
Docking2D & seed 2 & 0 & 6{,}894 & 3{,}889 & \multicolumn{3}{c}{no patch} \\
Docking2D & seed 3 & 3{,}309 & 6{,}878 & 3{,}866 & 5{,}249 & 3{,}362 & 1{,}166 \\
\bottomrule
\end{tabular}
\end{table}

\subsection{Patch Circuit}
\label{app:patch-size}
As described in Section~\ref{sec:patch-compilation}, a repair modifies a set of policy cells $\mathcal Q_r\subseteq\mathcal Q$, with each modified cell $q_r$ receiving an integer action correction $\Delta_r\in\mathbb Z^A$. The patch adds this correction to the controller's popcount whenever the input lies in a repaired cell, and leaves the output unchanged otherwise. We write $C:=|\mathcal Q_r|$ and number the repaired cells $q_1,\ldots,q_C$. Each is a product of one interval per input dimension, $q_r=\prod_{d=1}^{D}\{k_{r,d}\}$, where $k_{r,d}$ is its interval index along dimension $d$.

\paragraph{Guards.}
Let input dimension $d$ have $K_d$ thresholds $t_{d,1}<\dots<t_{d,K_d}$, inducing $K_d+1$ intervals, and thermometer bits
\[
    \tau_{d,j}(x):=\mathbf{1}[x_d>t_{d,j}],\qquad j=1,\ldots,K_d,
\]
so that interval $k$ of dimension $d$ is $(t_{d,k-1},t_{d,k}]$. Let $c_d(x)\in\{1,\ldots,K_d+1\}$ denote the index of the interval containing $x_d$. Then
\[
    c_d(x)=k\iff\tau_{d,k-1}(x)=1\ \wedge\ \tau_{d,k}(x)=0,
\]
where the first condition is omitted for $k=1$ and the second for $k=K_d+1$. Hence, membership in one interval is a function of at most two thermometer bits that the controller already computes. Each distinct interval is implemented once and shared between repaired cells, where they overlap. The guard of $q_r$ is the conjunction of its $D$ interval predicates $c_d(x)=k_{r,d}$, and it fires exactly for inputs in $q_r$. Since policy cells are disjoint, at most one guard fires.

\paragraph{Address encoder.}
The correction $\Delta_r$ is stored at address $r$. Address zero holds the zero correction and is selected when no guard fires. Since the guards are one-hot, bit $b$ of the $\lceil\log_2(C+1)\rceil$-bit address is the disjunction of the guards whose address has bit $b$ set.

\paragraph{Memory.}
The correction memory has $C+1$ words. Per action dimension $a$, it stores $(\Delta_r)_a$ in two's complement with the smallest width $v_a$ that represents $0$ and every $(\Delta_r)_a$. We synthesize the memory as a separate block and report its size separately.

\paragraph{Adders.}
For each action dimension $a$, the stored correction is sign-extended and added to the controller's popcount. Let $N_a$ denote the maximum popcount of action dimension $a$; the adder has the smallest two's-complement width $w_a$ that holds every sum, i.e.
\[
    -2^{w_a-1}\le\min\!\Big(0,\min_{r}(\Delta_r)_a\Big),
    \qquad
    N_a+\max\!\Big(0,\max_{r}(\Delta_r)_a\Big)\le 2^{w_a-1}-1 .
\]

\paragraph{Pipelining.}
The patch is a separate block next to the controller. Its guards, address and memory word are registered and the word is delayed, so the correction reaches the adder in the same cycle as the popcount of the same input. The patched controller thus keeps the latency of the original one. %

\end{document}